\documentclass[]{jfm} % dated: Oct '23
\usepackage{graphicx}
\usepackage{newtxtext}
\usepackage{newtxmath}
\usepackage{natbib}
\usepackage{hyperref}
\hypersetup{
    colorlinks = true,
    urlcolor   = blue,
    citecolor  = blue,
}

\newcommand{\RomanNumeralCaps}[1]
\usepackage{floatrow}
\usepackage[dvipsnames]{xcolor}
\usepackage{amssymb}
\usepackage{amsmath}
\usepackage{bm}
\usepackage{afterpage}
\usepackage[utf8]{inputenc}
\usepackage[position=top]{subfig}
\usepackage{caption}
\usepackage{booktabs}
\usepackage{float}
\usepackage{multirow}
\usepackage{mathtools,tikz}
\DeclareRobustCommand\sampleline[1]{%
  \tikz\draw[#1] (0,0) (0,\the\dimexpr\fontdimen22\textfont2\relax)
  -- (2em,\the\dimexpr\fontdimen22\textfont2\relax);%
}
\usepackage{url}
\definecolor{RoyalBlue}{rgb}{0.2549,0.4118,0.8824}
\definecolor{RYBGreen}{rgb}{0.2416,0.5749,0.1318} 

\usepackage{soul}
\usepackage{cleveref}
\usepackage{booktabs}
\title{Effect of rotation on wake vortices in stratified flow
} 
\author{Jinyuan Liu\aff{1}, Pranav Puthan\aff{1}, \and Sutanu Sarkar\aff{1,2,}\corresp{\email{ssarkar@ucsd.edu}} }
\affiliation{\aff{1}Mechanical and Aerospace Engineering, University of California San Diego, La Jolla, CA 92093, USA 
\aff{2}Scripps Institution of Oceanography, La Jolla, CA 92037, USA}

\begin{document}
\floatsetup[figure]{style=plain,subcapbesideposition=top}
\maketitle

%%%%%%%%%%%%%%%%%%%%%%%%%%%%%%%%%%%%%%%%%%%%%%%%%%%%%%
%%I have shortened it to be within the word count
\begin{abstract}
Stratified wakes past an isolated conical seamount are simulated at 
a Froude number of  $Fr = 0.15$ and  Rossby numbers of   $Ro = 0.15$, 0.75, and $\infty$.
 The wakes exhibit 
a K{\' a}rm{\' a}n vortex street,  unlike their unstratified, non-rotating counterpart. Vortex structures are studied in terms of large-scale global modes, as well as spatially localised vortex evolution, with a focus on  rotation effects. The global modes are extracted by spectral proper orthogonal decomposition (SPOD). For all three studied $Ro$ ranging from mesoscale, submesoscale, and non-rotating cases, the frequency of the SPOD modes at different heights remains coupled as a global constant.
However, the shape of the SPOD modes changes from slanted `tongues'
 at zero rotation ($Ro=\infty$) to tall hill-height columns at strong rotation ($Ro=0.15$). 
A novel method for vortex centre tracking 
 shows that, in all three cases, the vortices at different heights advect
uniformly at about $ 0.9U_{\infty}$ beyond the near wake, consistent with the lack of variability of the global modes. Under system rotation, cyclonic vortices (CVs) and anticyclonic vortices (AVs) present considerable asymmetry, especially at $Ro = 0.75$. The vorticity distribution as well as the stability of AVs are tracked downstream using statistics conditioned to the identified vortex centres. At $Ro=0.75$, intense AVs with relative vorticity up to $\omega_z/f_{\rm c}=-2.4$ are seen with small regions of instability  and they maintain large $\omega_z/f_{\rm c}$ magnitude in the far wake.  Recent stability analysis that 
  accounts for stratification and viscosity is found to improve on earlier criteria and show that these intense AVs are stable.
\end{abstract}

%%%%%%%%%%%%%%%%%%%%%%%%%%%%%%%%%%%%%%%%%%%%%%%%%%%%%%
% % \tableofcontents
%%%%%%%%%%%%%%%%%%%%%%%%%%%
\section{Introduction} \label{intro}
%%%%%%%%%%%%%%%%%%%%%%%%%%%

{The planet that we live on is full of multi-scale eddies, generated by various sources ranging from uneven thermal energy distribution at the largest scales, wind-driven ocean surface motions, to relatively more localised obstacle-induced flows. The dynamics of such eddies are greatly enriched by incorporating stratification, rotation, and turbulence. Understanding these dynamics is essential to geophysics. This work will be focused particularly on the last example -- wake eddies generated by obstacles. }

In the deep ocean, seamounts and hills are stirring rods; they induce vortical motion, turbulence, and internal gravity waves, which enhance heat and mass transport and hence crucially impact the ocean state. In the atmosphere, mountains commonly trigger wakes and waves, and orographic lifting is a source of convective weather, including air unsteadiness, formation of cumulonimbus clouds, and precipitation. Both the ocean and the atmosphere are stratified, and the scales of motions are large enough to feel the effect of Earth's rotation, leading to distinctive wake dynamics. The understanding of the wakes behind three-dimensional (3D) obstacles from a fluid dynamics perspective would benefit the modelling and prediction of the {multi-scale} motions of oceanic and atmospheric bottom boundary flows. 

% \afterpage{
\begin{figure}[t!]
  \centering
  \includegraphics[width=0.95\textwidth]{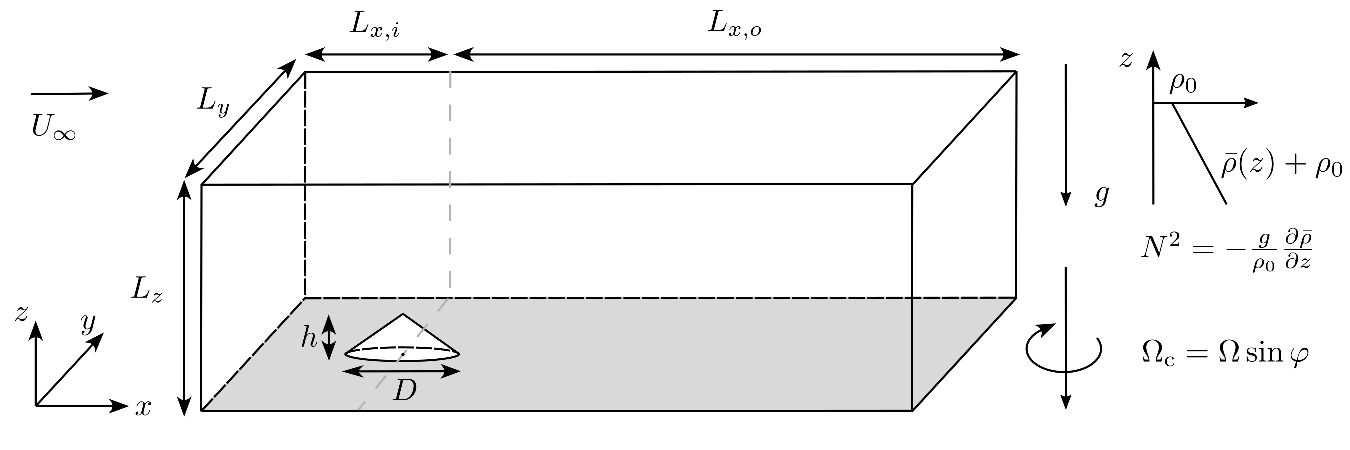}
  \caption{Flow configuration (not to scale). The background density is linearly stratified. The frame is rotating at a constant angular velocity $\Omega_{\rm c}=\Omega \sin \varphi$, where $\Omega$ is the Earth's angular velocity and $\varphi$ is the latitude. The Coriolis frequency, $f_{\rm c} = 2\Omega_{\rm c}$, is negative.  Gravity, density stratification and the axis of rotation all point to negative ${\bm{\hat{z}}}$.}
  \label{fig:flow} 
\end{figure}
% }

In this study, we consider the wake of a steady uni-directional mean flow ($U_\infty$)  {perturbed} by an isolated conical seamount/hill submerged in the fluid  (figure \ref{fig:flow}).   The background has stable density stratification that is linear so that the buoyancy
%Brunt-V{\" a}is{\" a}l{\" a} 
frequency $N=\sqrt{-g\partial_z \bar{\rho}/\rho_0}$ is a constant. The  Coriolis frequency $f_{\rm c}=2\Omega_{\rm c}$ is a negative constant (Southern hemisphere).   
%The direction of gravity, stratification, and rotation all point to the negative direction of the vertical axis. The geometry of the hill is a body of revolution with the rotating axis parallel to $z$. 
{A conical obstacle of base diameter $D$ and height $h$ is placed on a flat bottom wall to represent an idealized isolated hill/seamount.}
%It has base diameter $D$, height $h$, and a slope of approximately $30^\circ$ ($28.4^\circ$) such that $h/D = 0.3$.  
%with the center of the base circle at the origin. 

Two non-dimensional numbers, the vertical Froude number ($Fr=U_{\infty}/N h$) and the Rossby number ($Ro=U_{\infty}/|f_{\rm c}| D$) are the main controlling parameters. 
%Here $h$ and $D$ are the vertical and horizontal dimensions of the hill, respectively.
An additional (but not independent)  non-dimensional number, the Burger number ($Bu=N^2h^2/f_{\rm c}^2 D^2 =(Ro/Fr)^2$) that characterises the  importance of  stratification relative to rotation will also be used. {The dynamical significance of the Burger number can also be interpreted as the normalised Rossby radius of deformation, $L_d/D=Nh/f_{\rm c}D = \sqrt{Bu}$.} 
%\rev{The slope of the obstcle, $h/D$, will also play a crucial role in the dynamics. It is fixed to about $30^\circ$ in this study to represent flow past a steep (with respect to internal wave propagation) obstacle instead of along a shallow continental slope.}

Stratified hill wakes have been studied extensively through laboratory experiments, field observations, and numerical simulations. \cite{hunt1980experiments}  showed experimentally, that for relatively strong stratification ($Fr < 0.4$),  there is a potential energy barrier below which the flow does not have sufficient  kinetic energy to go over the hill and, instead, flows around the hill to form a quasi-two-dimensional (Q2D) K{\' a}rm{\' a}n vortex shedding (VS) pattern. This is a significant qualitative difference between stratified and unstratified flow past a three-dimensional obstacle. Without stratification, the flow goes over the obstacle and  horseshoe structures are formed \citep{garcia2009large}. \cite{castro1983stratified} showed in stratified flow past a finite-span ridge that, as $Fr$  decreases from above to below unity, the wake behind the hill transitions from a lee-wave-dominated regime to a vortex-dominated regime, with the latter regime being the focus of this work. They found that at or below $Fr=0.2$ (their figure 7, last row), the modulation of the vortex wake by the lee waves can be neglected except near the peak of the ridge. \cite{boyer1987stratified} studied experimentally the effect of system rotation on stratified hill wakes in the regime of $Ro = O(0.1)$, $Fr = O(0.1)$-$O(1)$, equivalently $Bu = O(1)$-$O(10)$ variation. They found that the VS frequency is not strongly affected by changing  $Ro$. However, the Reynolds number ($Re_D<1200$) was not sufficient for the wake to be turbulent. For non-rotating hill wakes, \cite{vosper1999experimental} varied the Froude number and the shape of the wake generator, and found that the VS frequency is a weak increasing function of $Fr$, and the $St$-$Fr$ relationships collapses for different object shapes if the reference velocity is corrected for blockage. More recently, \cite{teinturier2010small,lazar2013inertialb} performed laboratory experiments on the LEGI-Coriolis rotating platform where a cylinder ($h/D$ = 0.1 was small) was towed in the upper layer of a fluid with two-layer stratification as a simple model of island wakes. The towing velocity and rotation speed were varied to achieve different combinations of $Ro$ and $Re$ with constant $Re/Ro$, e.g. $Ro =1$ corresponded to $Re = 10,000$ in \cite{teinturier2010small} and the experiments of \cite{lazar2013inertialb} considered $Re = 2000$-$7000$. With these laboratory experiments, they were able to investigate the asymmetry between cyclonic and anticyclonic vortices over a range of $Ro$.

 %at moderate-to-high Reynolds number ($Re_D \sim O(10^{3}) {\text -} O(10^{4})$) that featured the asymmetry between cyclonic and anticyclonic vortices generated behind a towed cylinder ($h/D$ = 0.1 was small).  
 
 %On the other hand, doing numerical simulations allows one to change the flow conditions easily and to vary one parameter at a time to study its effect individually. 
For rotating stratified hill wakes,  numerical studies include the utilization of the hydrostatic regional oceanic modelling system (ROMS) \citep{shchepetkin2005regional}, such as \cite{dong2007island,perfect2018vortex,perfect2019energetics,srinivasan2021high,jagannathan2021boundary}, and the {hydrostatic version} of 
the MIT-GCM model \citep{marshall1997hydrostatic,marshall1997finite}, such as \cite{liu2018numerical}. 
%The  vertical momentum equation is commonly simplified  into a hydrostatic balance between vertical pressure gradient and gravity), hence the Poisson equation solving the pressure is reduced to a two-dimensional equation which can be solved at a much lower cost. This assumption could be correct to at least the leading order but not the entire flow regime and parameter space. A more detail discussion of this assumption and a \textit{posteriori} analysis will be given in Appendix A. 
Recently, \cite{puthan2021tidal,puthan2022high,puthan2022wake} conducted {non-hydrostatic} large-eddy simulation (LES) of stratified hill wakes.
% without the hydrostatic approximation.
% to capture not only  wake vortices but also turbulent dissipation. 
%\jy{ A similar numerical procedure will be followed in this work.} 
{The focus was on the role of tidal modulation of a mean current in a regime with strong stratification but weak rotational effects. The findings included tidal synchronization  (vortex shedding at specific tidal subharmonics that  depend on the tidal excursion number), phases with enhanced turbulent dissipation, and higher form drag.  A similar numerical procedure will be followed in this work while excluding the periodic tide and focussing on rotation effects. }

In terms of field observations, island wakes have acquired much attention recently and wakes have been studied near Palau \citep{mackinnon2019eddy,zeiden2021broadband}, the Green Island off the coast of Taiwan \citep{chang2019observations}, and in the lee of Guadalupe \citep{horvath2020evolution}, to name a few. {A strong sign of a narrow-band VS mode stood out from the broadband turbulence signal, and coherent submesoscale vortices were found in the wake. Submesoscale vortices are characterised by a Rossby number of order unity, and are receiving increasing attention \citep{taylor2023submesoscale}.
%Those observations were made in \textit{in situ} conditions that have the complications of  irregular terrain, tidal motion, and nonlinear stratification, but the amount of data -- both in space and in time -- is insufficient for a more comprehensive study (possible using simulations) of the spatio-temporal structure of the wake.
}
%This limitation along with the possibility of a controlled parametric study of coherent wake vortices motivates our high-fidelity numerical simulations in setting of a topographic wake.} 

In this paper, we  present results from seamount/hill wakes at $Fr = 0.15$ and three representative Rossby numbers, $Ro=0.15, 0.75, \infty$, corresponding to mesoscale, submesoscale, and non-rotating stratified flow regimes, respectively.  The conical hill has $h/D = 0.3$ and the Reynolds number is $Re = 10,000$.  {There is near-wake turbulence but the focus of this paper will be on characterising coherent wake vortices that emerge further downstream.} 

{The preceding literature survey raises several scientific issues regarding wakes of isolated topography in a rotating, stratified fluid. Previous simulations of an obstacle wake with significant stratification and rotation effects have used the hydrostatic assumption. In contrast,  our non-hydrostatic model and LES approach (with high resolution as will be described) enables the capture of three-dimensional turbulent motions and, additionally,   topographic internal gravity waves. The temporal frequency of vortex shedding (VS) from a conical obstacle in stratified flow can depend on the height of measurement at the obstacle in addition to background stratification and rotation. Current understanding of the VS frequency at a conical obstacle, which is based on a laboratory study~\citep{boyer1987stratified} and hydrostatic simulations~\citep{perfect2018vortex}, as well as the spatial structure of the wake vortices is incomplete. With the application of spectral proper orthogonal decomposition  (section \ref{large-scales}) to vertical vorticity, we are able to not only pinpoint the VS frequency and its dependence on $Ro$ ($Bu$) but also the spatial organization of the associated coherent structure. Another question is the relation with linear stability theory of the stable coherent wake eddies (their vorticity and radius) that emerge in the simulations from the strongly turbulent near wake.  A novel eddy identification scheme allows  statistical measurement, based on a large ensemble of individual realizations, of the wake-eddy structure   and its evolution as a function of downstream distance (section \ref{tracking}). Our comparison with stability theory (section \ref{cyclones}) expands on the important laboratory findings of \cite{lazar2013inertialb} in various ways: substantially higher $Re = 10,000$ at $Ro = 0.15$ relative to the laboratory value of $Re = 1500$, the non-uniform cross-section of a cone instead of a cylinder, and a continuous linear stratification instead of a two-layer stratification.
}

The rest of the paper is structured as follows: Section \ref{simulation} introduces the numerical methods and the LES simulations conducted  as part of  this work; section \ref{large-scales} elucidates the global structures in the flow that are coherent in space and time via flow visualisations and spectral proper orthogonal decomposition (SPOD); section \ref{tracking} presents  ensemble-averaged tracks of  the centres of vortices that are obtained by application of a pattern recognition method; %reveal the statistics of the intensity and advection velocities during their evolution; 
section \ref{cyclones} is a systematic study of the asymmetry between 
% \st{and instabilities of cyclones and anticyclones} 
{cyclones and anticyclones and the instabilities of the anticyclones}; section \ref{statistics} concerns the evolution of the mean momentum wake; 
%analyses the mean and turbulent statistics of the bulk of flow and conditioned to extracted vortex centres;
 and, finally, section \ref{conclusion} is a summary and discussion of the results. 
 
%  \Sutanu{In last section, highlight the important results and discuss these results in context of what is known.}
 %makes an overall review and concludes the entire paper. 

% Motivation of this work is that the the information about the spatio-temporal coherence of the structures is under investigated because of lack of data. We found a global mode. People did look at the spatial (plots) and temporal (one point spectra) of this flow but none combined space and time.   

  %%%%%%%%%%%%%%%%%%%%%%%%%%%%%%%%%%%%%%%%%%%%%%%%%%%%% 
\section{Numerical simulations} \label{simulation}

% We solve  as in Eq. \ref{equations}. 

The incompressible Navier-Stokes equations are solved in a Cartesian coordinate with $x,y,$ and $z$ being the streamwise, transverse, and vertical directions, as shown in figure \ref{fig:flow}.  In the momentum equation, density variation appears only in the buoyancy as per the Boussinesq approximation and system rotation is represented by the Coriolis force. {The non-dimensional governing equations in index notation are as follows:} 
{
% \begin{align}
%   \frac{\partial u_i}{\partial x_i} & = 0, \\ 
%   \frac{\partial u_i}{\partial t} + \frac{\partial u_i u_j}{\partial x_j} -  f_{\rm c} \epsilon_{ij3} (u_j - U_{\infty}\delta_{j1}) & = -\frac{1}{\rho_0} \frac{\partial p^*}{\partial x_i} 
%   + \frac{\partial \tau_{ij}}{\partial x_j} 
%   - \frac{\rho^* g }{\rho_0} \delta_{i3}, \label{eqn:momemtum} \\ 
%   \frac{\partial \rho}{\partial t} 
%   + \frac{\partial \rho u_i}{\partial x_i}  & = 
%   \frac{\partial  {{J}_{\rho,i}}}{\partial x_i}, \\
%   \tau_{ij} = (\nu + \nu_{\rm sgs}) (\frac{\partial u_i}{\partial x_j} + \frac{\partial u_j}{\partial x_i}) , \, 
%    {J_{\rho,i}} &= (\kappa + \kappa_{\rm sgs})\frac{\partial \rho}{\partial x_i}.  
%   \label{equations}
% \end{align} 
\begin{align}
  \frac{\partial u_i}{\partial x_i} & = 0, \\ 
  \frac{\partial u_i}{\partial t} + \frac{\partial u_i u_j}{\partial x_j} -  \frac{1}{Ro} \epsilon_{ij3} (u_j - U_{\infty}\delta_{j1}) & = -  \frac{\partial p^*}{\partial x_i} 
  + \frac{\partial \tau_{ij}}{\partial x_j} 
  - \frac{1}{Fr_D^2} \rho^*\delta_{i3}, \label{eqn:momemtum} \\ 
  \frac{\partial \rho}{\partial t} 
  + \frac{\partial \rho u_i}{\partial x_i}  & = 
  \frac{\partial  {{J}_{\rho,i}}}{\partial x_i}, \\
  \tau_{ij} = \frac{1}{Re} \left(1 + \frac{\nu_{\rm sgs}}{\nu} \right) (  \frac{\partial u_j}{\partial x_i} + \frac{\partial u_i}{\partial x_j}) , \, 
   {J_{\rho,i}} &= \frac{1}{RePr} \left(1 + \frac{\kappa_{\rm sgs}}{\kappa} \right)\frac{\partial \rho}{\partial x_i},   
  \label{equations}
\end{align} 
} {where the normalising units for $x_i$, $u_i$, $t$, $p^*$, and $\rho^*$ are $D$, $U_{\infty}$, $D/U_{\infty}$, $\rho_0 U_{\infty}^2$, and $-\partial_z \bar{\rho} D$, respectively. Here the horizontal Froude number, $Fr_D=U_{\infty}/ND$, is related to the vertical Froude number used in this study as $Fr_D = (h/D) U_{\infty}/Nh = (h/D) Fr = 0.045 $. The Prandtl number, $Pr=\nu/\kappa$, which is the ratio between the molecular viscosity $\nu$ and the scalar diffusivity $\kappa$, is set to unity.}  

The total density $\rho$ is decomposed into the reference density $\rho_0$, the linearly varying background density  $\bar{\rho} (z)$, and the density perturbation $\rho^*$ due to fluid motion, 
\begin{equation}
  \rho(x,y,z,t) = \rho_0 + \bar{\rho}(z) + \rho^*(x,y,z,t). \label{rho_a}
\end{equation} 
The total pressure is written as
\begin{equation}
  p(x,y,z,t) = p_0 + p_g(y) + p_a(z) + p^*(x,y,z,t),  \label{p_a}
\end{equation} where  {the reference pressure $p_0$ is a constant,} the hydrostatic (ambient) pressure $p_a$ has a vertical gradient that balances the ambient density ($\rho_a=\rho_0 + \bar{\rho}(z)$), and the geostrophic pressure $p_g$ has a transverse gradient that balances the Coriolis force due to the velocity $U_{\infty}$ of the freestream.  Only the dynamic pressure $p^*$ appears in the momentum equation \eqref{eqn:momemtum}. 
%The molecular and subgrid-scale kinematic viscosity $\nu, \nu_{\rm sgs}$ are defined based on $\rho_0$. 

The LES is performed at a  {moderately high} Reynolds number $Re_D = U_{\infty} D/\nu = 10 \, 000$. Spatial derivatives are discretized with a second-order central finite difference on a staggered grid, and the equations are advanced in time using a combined scheme with  third-order Runge-Kutta for the convection terms and Crank-Nicolson for the diffusion terms. Continuity is enforced by solving the pressure Poisson equation. The obstacle is represented by an immersed boundary method  \citep{balaras2004modeling,yang2006embedded}. The sub-grid-scale (SGS) model is chosen to be the WALE model \citep{nicoud1999subgrid} with the SGS Prandtl number $Pr_{{\rm sgs}}=\nu_{\rm sgs}/\kappa_{\rm sgs}$ set to unity, {where the SGS viscosity $\nu_{\rm sgs}$ and diffusivity $\kappa_{\rm sgs}$ represent the modelled effects on the transport of resolved momentum and scalar by the unresolved scales.} For more numerical details, the reader is referred to \cite{puthan2020wake}. 

{The conical obstacle has base diameter $D$, height $h$, and a slope of approximately $30^\circ$ ($28.4^\circ$) such that $h/D = 0.3$. The slope of about $30^\circ$ in this study is steep in the oceanic context,  both dynamically (much larger than typical internal wave propagation angles) and geometrically (typically, underwater bathymetry has much smaller slope angle).}

The computational domain spans a volume of $L_x \times L_y \times L_z = [-4,15] \times [-4,4] \times [0, {4}]$  {in units of $D$} and the horizontal resolution of the immersed body and the turbulent near wake ($-1<x/D<2$) is held constant at $(\Delta x, \Delta y)\approxeq(0.003D,0.006D)  {\le} (4\eta, 8 \eta)$, with a  mild stretching in the streamwise direction. Here $\eta$ is the minimum Kolmogorov length scale at the centerline at different heights. The vertical resolution below $z/h=1.2$ is kept at  {$\Delta z = 0.008h \approxeq 0.05 U_{\infty}/N$} to resolve the length scale for vertical overturning motion $U_{\infty}/N$.
% is properly resolved as well as the vertical mean shear $\partial_z \langle U \rangle$. 

The inflow condition is a uniform velocity inlet, and the outflow is a Neumann-type convective outlet. The lateral boundaries are periodic to reduce the blockage effect and to allow the wake to flap. {The top boundary is shear free, and the hill boundary is no-slip for velocity and no-flux for density.} 
% The top boundary is shear free, and the bottom boundary is modelled by a quadratic drag law as in \cite{puthan2020wake,jagannathan2021boundary}. The drag coefficient is chosen as $c_D=0.002$ in common with oceanography applications \citep{haidvogel1999numerical,arbic2008quadratic}. Its value was validated in a stratified bottom Ekman layer by \cite{taylor2008stratification} and tested in stratified flow over complex geometry by \cite{rapaka2016immersed}, among others. 
Sponge layers are placed at the inlet and the top boundaries to reduce spurious reflected internal wakes. 
% \st{The overall numerical setting was validated by}
{The overall LES setting and the immersed boundary formulation have been  tested and validated  against available data on force coefficients and evolution of wake deficit and turbulence intensity for flow past various geometries (sphere, disk, cone) in unstratified and stratified environments} \citep{pal2016regeneration,pal2017direct,chongsiripinyo2020decay,puthan2020wake}. 
% In this particular setting of rotating stratified hill wakes, \cite{perfect2019energetics} had examined that varying the drag coefficient does not qualitatively change the characteristics of VS in the wake and has little influence above the bottom boundary layer, which is estimated to be below $0.1h$ in our cases. Hence, analysis of the resulting vortex wake will only be conducted for locations at least $0.1h$ from the bottom wall to exclude the effect of the modeled bottom boundary and the turbulence therein. The first layer of grid on the solid wall of the hill is also modeled with the same drag law and coefficient for consistancy and the sake of eliminating the grid requirement for resolving the boundary layer on the object.  

The parameter space explored is shown in table \ref{table1}. The stratification  is held constant at $Fr = 0.15$,   a typical value for midsize topography in the ocean and the atmosphere, and is in the `flow-around' regime where coherent wake vortices dominate. %\citep{hunt1980experiments}.
Meanwhile, the rotation Rossby number is varied as $Ro=0.15, 0.75, \infty$, to study its effect on the vortex wake. These three values of $Ro$ correspond to mesoscale, submesoscale, and non-rotating geophysical flows. The induced Burger number are $Bu = 1, 25, \infty$, which will be used to label different cases. 

\begin{table}[t!]
  \caption{Parameters of simulated cases. $N_x,N_y,$ and $N_z$ are the number of grid points in each direction and $N_t$ is the total number of available snapshots that will be used for all statistics. 
  %  and the SPOD analysis in Section \ref{large-scales}. 
   $ T$ is the time span of the stored data. The aspect ratio of the conical obstacle is $h/D = 0.3$ for all cases.}
  \begin{tabular}{l c c c c c c c}
    % \hline 
    \toprule
    Case &$Bu$&$Ro$ & $Fr$                  & $(N_x,N_y,N_z)$ & $N_t$ & $T U_{\infty}/D$ & color code \\ 
    % \hline 
    \midrule 
    BuInf  &$\infty$   &$\infty$& \multirow{3}{*}{0.15} & \multirow{3}{*}{(1536,1280,320)} &  \multirow{3}{*}{4000} & 295  & red\\ 
    Bu25 & 25       & 0.75      &          &  & & 345 & green\\
    Bu1  & 1 & 0.15 &          &  & & 322 & blue \\
    % \hline 
    \bottomrule 
  \end{tabular} 
  \label{table1}
\end{table} 

We compile a time-resolved numerical database that consists of three rotation strengths and collect the data after statistical stationarity is reached. Each case has $N_t=4000$ snapshots that span around 300 convective time units ($T = 300 D/U_{\infty}$), which corresponds to roughly $80$ VS periods. Statistics, to be discussed later,  are obtained by averaging  over the entire interval, $T$.  

  %%%%%%%%%%%%%%%%%%%%%%%%%%%
\section{Large-scale coherent structures}   
\label{large-scales}
%%%%%%%%%%%%%%%%%%%%%%%%%%%
In geophysical flows, coherent vortical structures are commonly observed. 
%The large scales are determined by the external boundaries and are typically proportional to the size of eddy generators that are kilometers wide in nature. The resulting lifetimes of those structures are also long, signifying their environmental and engineering implications. Coherent structures are organised motions in turbulent flows. 
There is no universal definition of coherent structures, but they are generally strong enough and have relatively independent dynamics to distinguish themselves from the background flow, account for a significant  portion of the fluctuation energy of the system, are spatially organized, and their lifetime is sufficient  for them to be dynamically important. The  wake eddies of this paper have  the aforementioned features.

Large-scale K{\' a}rm{\' a}n vortices, a specific type of coherent structure, are associated with vortex shedding from bluff bodies. In geophysical applications, {they are commonly found in  flows  impinging on bottom or side topography, e.g.   island wakes } \citep{young2006observational,chang2019observations,horvath2020evolution}, headland wakes \citep{pawlak2003observations}, and in laboratory flows  \citep{hunt1980experiments,castro1983stratified,boyer1987stratified,teinturier2010small}, among others. In field observations and laboratory experiments, the features of the coherent vortices are inferred from single- or multiple-point statistics and flow visualizations. The data is limited in spatial coverage and resolution.

In what follows, coherent structures will be studied in two ways. First,   individual snapshots in which vortex structures are vividly visible are used for a qualitative overview  of rotation effects  (section \ref{viz}).  Second, a more comprehensive quantitative assessment  is obtained by applying  SPOD to  the time-resolved LES database in table \ref{table1} to reveal the statistical significance of the coherent structures. Section \ref{spod-theory} reviews the theory and implementation of SPOD, followed by the analysis of the temporal eigenspectra (section \ref{eigenspectra}) and spatial modes (section \ref{eigenmodes}). 

% \Sutanu{ Jinyuan, we will have to justify the use of $\omega_z$, excluding other quantities. Text as follows ...}

%\jy{jy: I've combined your text and my justification in a latter section in the SPOD part and put it there.}

% In the cases with strong rotation, the vertical vorticity of  is of particular interest as it adjusts toward a more stable state, which will be examined in detail in section \ref{cyclones}. 
%
%\st{Coherent dynamics will be studied in a region $x > ??$ that excludes near-wake turbulence and where wake vorticity is dominated by $\omega_z$ as shown by figure ?? (a) that compares the three vorticity components on a horizontal plane.   The velocity is dominated by the horizontal components as shown in figure ?? (b). Thus, the stratified wakes are quasi-two dimensional insofar as the velocity components, although there is vertical shear between adjacent planes. Since our interest lies in the coherent structures of  horizontal motion,  the visualization and SPOD analysis are presented for the vertical relative vorticity $\omega_z$. There is no need to employ criteria based on $Q$ or $\lambda_2$ that are necessary in turbulent flows.  }  
%
%\jy{The spanwise vorticity seems not to be small and I could probably not say $\omega_z$ dominates vorticity. }

\subsection{Flow visualisations} \label{viz}

\begin{figure}[t!]
  \captionsetup[subfloat]{farskip=-24pt,captionskip=1pt,labelformat=empty}
  \subfloat[]{{\includegraphics[width=.85\linewidth]{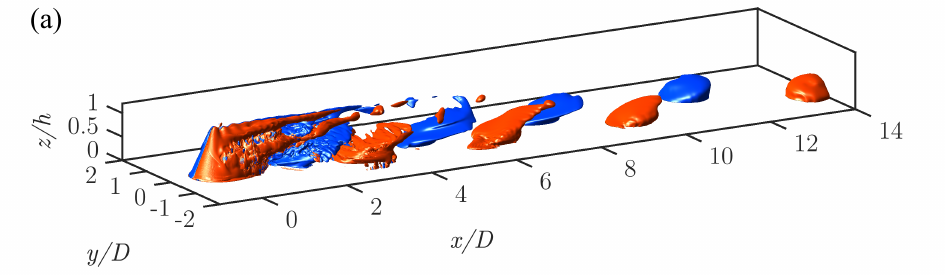}}}% 
  \label{fig:vis-Buinf}
  \subfloat[]{{\includegraphics[width=.85\linewidth]{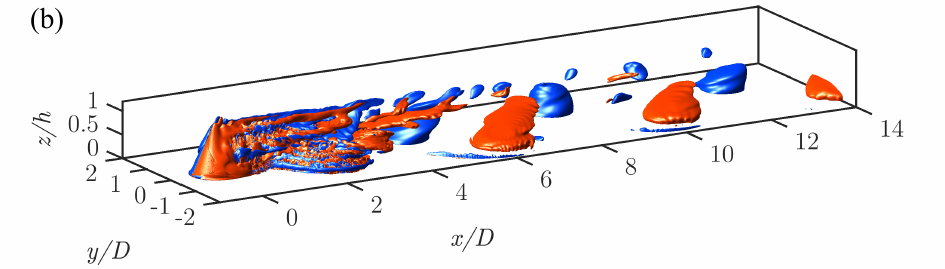}}}% 
  \label{fig:vis-Bu25} \\ 
  \subfloat[]{{\includegraphics[width=.85\linewidth]{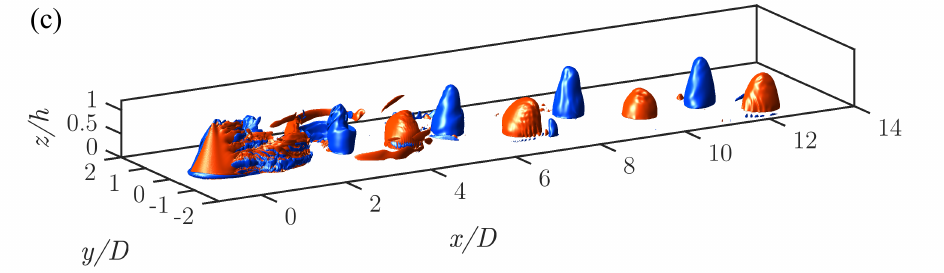}}}% 
  \label{fig:vis-Bu1}  
  \caption{Visualisation of the isosurfaces of $| \omega_z | D/U_{\infty}=1.5, 1.0, 3.0$ for BuInf, Bu25, and Bu1 (top to bottom), respectively. Red and blue indicate positive (anticyclonic since $f_{\rm c}$ is negative) and negative (cyclonic) vorticity, respectively. Note the vertical axis is normalised with the height $h$ of the hill, which is about $0.3$ times the base diameter $D$. {The three snapshots (a-c) are taken at time instants $tU_{\infty}/D=252.0, 221.6, 152.0$, respectively.}}
  \label{fig:3dvis} 
\end{figure}

A first impression of the 3D spatial organization of the wake vortices is provided by the  isosurfaces of the vertical component of vorticity  {($\omega_z=(\nabla \times \bm{u})_z$)} in  figure \ref{fig:3dvis}(a-c) for cases BuInf, Bu25, and Bu1, respectively. {At moderately strong stratification ($Fr\le O(0.1)$), vertical overturning motions are surpressed to approximately one order of magnitude smaller than horizontal flows, with the latter well-represented by $\omega_z$.} 

\begin{figure}[t!] 
  \captionsetup[subfloat]{farskip=-6pt,captionskip=1pt}
  \subfloat[]{{\includegraphics[width=.45\linewidth]{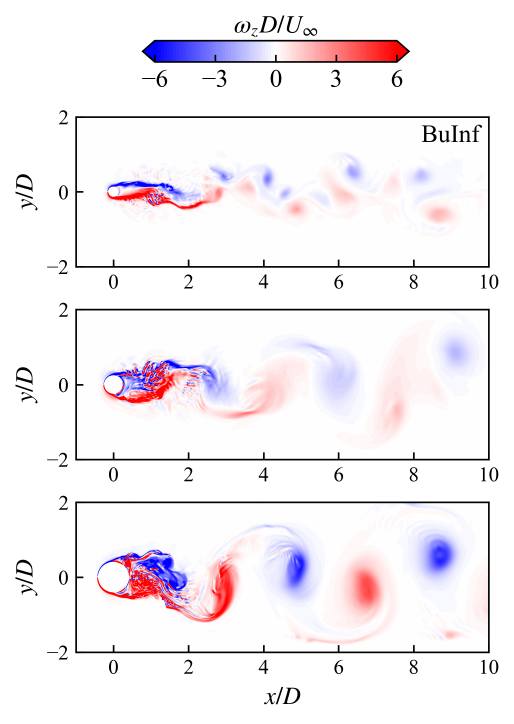}}}% 
  \subfloat[]{{\includegraphics[width=.45\linewidth]{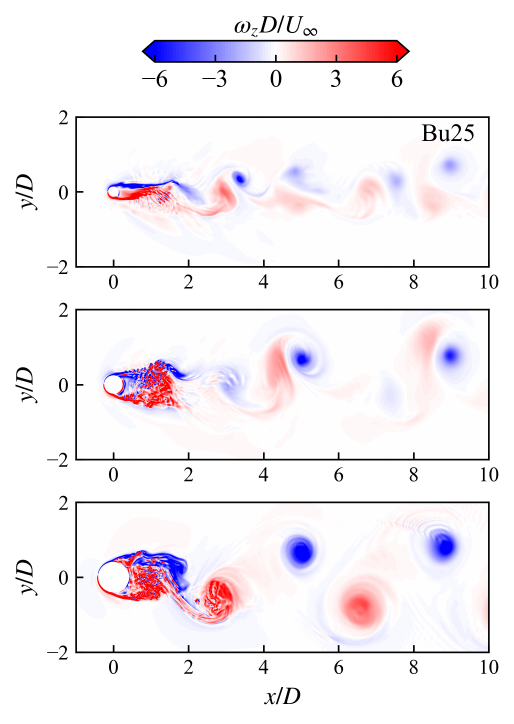}}}% 
  \\ 
  \subfloat[]{{\includegraphics[width=.45\linewidth]{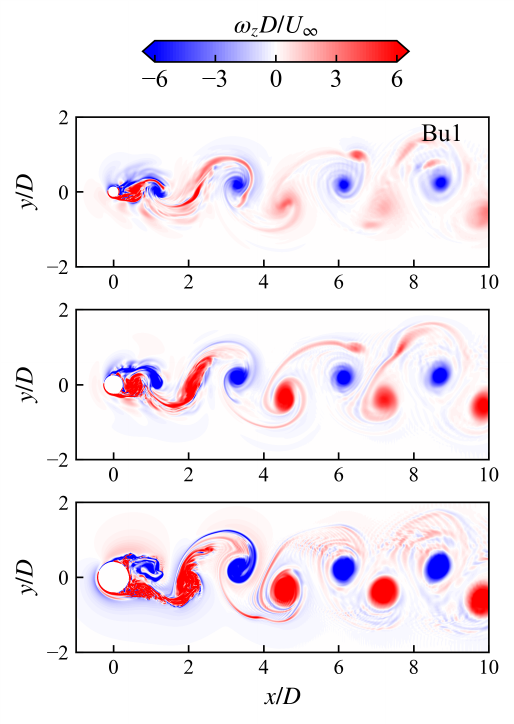}}}% 
  \subfloat[]{{\includegraphics[width=.45\linewidth]{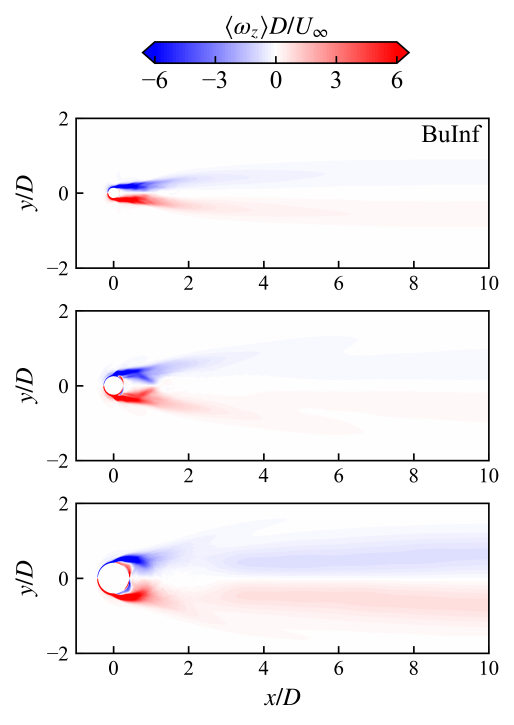}}}% 
  \caption{(a,b,c) Instantaneous and (d) time-averaged $\omega_z$ at horizontal planes at $z/h=0.75, 0.50, 0.12$ (from top to bottom in each subfigure). (a,d) Cases BuInf, (b) Bu25, (c) Bu1. {In (a-c), the snapshots are taken at non-dimensional time instants $tU_{\infty}D= 77.9, 123.9,63.3$ counted after the fully developed state is reached, respectively.}}
  \label{fig:omgz_mean}
\end{figure} 

%It is shown in figure \ref{fig:3dvis}(a-c) that 
In the near-wake region ($x/D<3$), the $\omega_z$ isosurface is space-filling and multiscale,   indicating greater turbulence intensity than further downstream. But after $x/D=3$, the vortices quickly organise into spatially compact coherent structures that persist to  the end of the computational domain with little change.
% during their advection. 
 Rotation clearly influences the shape and size of the structures. For case BuInf in figure \ref{fig:3dvis}(a), the structures are slanted {`tongues'} with  horizontal dimensions greater than their  height. As the strength of rotation is increased, the horizontal size shrinks   and the height increases  as in figure \ref{fig:3dvis}(b) for Bu25. With further increase of rotation,  the vortex structures become aligned with the vertical axis, appearing as  tall `columns'  that extend from the flat bottom to the hill height, in figure \ref{fig:3dvis}(c). 

The wake vortices are composed of cyclones (rotation is in the same direction as that of the frame) and anticyclones. It is  found in figure \ref{fig:3dvis}(b,c) that, with the presence of rotation, the cyclones (negative vortices in blue) are different from anticyclones (positive vortices in red) with regards to size, shape, and vorticity distribution as will be elaborated in later sections. In figure \ref{fig:3dvis}(c), cyclones are taller and thinner than anticyclones and, as will be shown,  have stronger interior vorticity.

Figure \ref{fig:omgz_mean}(a,b,c) show the instantaneous $\omega_z$ at several heights for all three cases. In all three cases,  a pattern of  Q2D K{\' a}rm{\' a}n vortices is distinct { in  all planes at $z/h=0.12, 0.25, 0.75$. }
%  The heights in text are different from in the caption! and I see a Karman wake at all 3 planes}.
 The flow is akin to a vortex wake rather than its unstratified counterpart of a three-dimensional turbulent wake \citep{garcia2009large}. 

The mean vertical vorticity ($\langle \omega_z \rangle$) of case BuInf is shown in figure \ref{fig:omgz_mean}(d). For the lower two planes, the mean looks very similar to those obtained in low Reynolds number two-dimensional cylinder wakes,  {such as} in \cite{barkley2006linear,mittal2008global}, both at $Re_D=100$. Such similarity supports the Q2D feature of the hill wake at $Fr=0.15$. At the same time, there is a notable difference: the flow in each horizontal place that cuts the hill {\em does not} represent an independent two-dimensional  {(2D)} flow around a cylinder with the local hill diameter. Among different planes at different heights in figure \ref{fig:omgz_mean}(d), the length of the attached shear layer (with dark colors), is approximately a constant, regardless of the variation of the local hill diameter. This is consistent with the fact that the VS frequency is a {\em global constant} which will be discussed in detail in the next section, as the shear layer length is correlated to the shedding frequency \citep{williamson1998series}. 

{In the SPOD analysis of the next section, the vertical vorticity on different horizontal planes ($\omega_z(x,y,t;z)$) is chosen as the quantity of interest since the large-scale wake structures involve predominantly {Q2D vortical motion} (figures \ref{fig:3dvis}-\ref{fig:omgz_mean}), the shedding and evolution processes of which are well represented by $\omega_z$. Moreover, the focus of this work is on the influence of increasing rotation strength on the horizontal motions, which tends to constrain the flow to be around the vertical axis at the large scales and significantly enhance $\omega_z$ as will be shown later. In the rotating cases Bu25 and Bu1, the stability of the anticyclones will also be studied (section \ref{cyclones}) with $\omega_z$ being one of the most important metrics, hence we apply $\omega_z$ rather than other vortex identification criteria for overall consistency. }

{As seen in figures \ref{fig:3dvis}-\ref{fig:omgz_mean}, $\omega_z$ structures exhibit  dissimilar vertical organisation at different levels of rotation, although K{\' a}rm{\' a}n vortices are present in each horizontal plane. Owing  to stratification, a vertical length scale of $U_{\infty}/N = Frh = 0.15h$ is introduced to the flow, and whether  vortex structures  remain coherent over vertical distances larger than  $U_{\infty}/N$  {regardless of rotation} (suggested  {affirmatively} by figure \ref{fig:3dvis})  needs quantitative investigation. To that end, we perform SPOD on vertically offset horizontal planes at $z/h=0.12, 0.25,0.50,0.75$ ($N_x\times N_y\times N_z=1538\times 1280\times 1$) and the vertical center plane at $y=0$ ($N_x\times N_y \times N_z=1538\times 1\times 320$) to allow the choice of different dominant (vortex shedding) frequencies at different heights by the flow and avoid imposing {\it{a priori}} a global frequency.}

%%%%%%%%%%%%%%%%%%%%%%%%%%%%%%%%%%%%%%%%%%%%%%%%%%%%%%
\subsection{Spectral POD and its numerical implementation} \label{spod-theory} 
Proper orthogonal decomposition (POD) is a matrix-factorization-based modal decomposition of complex systems introduced into fluid mechanics by \cite{bakewell1967viscous,lumley1967structure,lumley1970stochastic}. 

Consider a statistically stationary square-integrable multi-variable signal $\bm{q}(\bm{x},t) \in \mathcal{L}_{\boldsymbol{\mathsfbi{W}}}^2(\bm{\Omega})$ 
with zero mean. Here $\mathcal{L}^2_{\boldsymbol{\mathsfbi{W}}}$ is the Hilbert space equipped with a weighted inner product \begin{equation}
  (\bm{q}_1, \bm{q}_2)_{\boldsymbol{\mathsfbi{W}}} = \int_{\bm{\Omega}} \bm{q}_2^{\rm H}(\bm{x})  \boldsymbol{\mathsfbi{W}}(\bm{x}) \bm{q}_1(\bm{x}) \, {\rm d} \bm{x} 
  \label{eqn:inner}
\end{equation} on a bounded domain $\bm{\Omega}$, and $(\cdot)^{\rm H}$ denotes Hermitian transpose. The weight matrix $\boldsymbol{\mathsfbi{W}}$ is Hermitian positive-definite and the weighted $2$-norm is defined as $\|\bm{q}_1\|_{\boldsymbol{\mathsfbi{W}}}=(\bm{q}_1,\bm{q}_1)_{\boldsymbol{\mathsfbi{W}}}^{1/2}$. The symbol $\langle \cdot \rangle_{\rm E}$ denotes the ensemble average over all realizations, and it is equivalent to the time average under ergodicity. The goal of POD is to find an empirical function $\boldsymbol{\psi}(\bm{x})$ that solves the optimization problem \begin{equation}
  \boldsymbol{\psi}(\bm{x}) = \arg \max_{\|\boldsymbol{\psi}\|_{\boldsymbol{\mathsfbi{W}}}=1} \langle (\bm{q},\boldsymbol{\psi})_{\boldsymbol{\mathsfbi{W}}} \rangle_{\rm E}, \label{eqn:ops}
\end{equation} which defines $\boldsymbol{\psi}(\bm{x})$ as the function on which the projection of $\bm{q}(\bm{x},t)$ is maximized in the sense of the least squares. Since $\mathcal{L}^2_{\boldsymbol{\mathsfbi{W}}}$ is an infinite-dimensional space, a practical way to obtain the empirical function $\boldsymbol{\psi}(\bm{x})$ is to approximate it within a finite-dimensional space spanned by $\{\boldsymbol{\psi}^{(i)}\}_{i=1}^{M}$, where $\boldsymbol{\psi}^{(i)}$ is the $i$-th spatial mode and $M$ is the order of truncation. It was shown in \cite{holmes2012turbulence} that the optimization problem \eqref{eqn:ops} can be converted to a Fredholm eigenvalue problem as \begin{equation}
  \mathcal{R} \boldsymbol{\psi}^{(i)}(\bm{x})  = \int_{\bm{\Omega}} \boldsymbol{\mathsfbi{R}}(\bm{x},\bm{x}')\boldsymbol{\mathsfbi{W}}(\bm{x}') \boldsymbol{\psi}^{(i)}(\bm{x}')  \, {\rm d} \bm{x}' = \lambda^{(i)} \boldsymbol{\psi}^{(i)}(\bm{x}) , 
  \label{eqn:evp} 
\end{equation} where $\mathcal{R}$ is a linear operator and $\boldsymbol{\mathsfbi{R}}(\bm{x},\bm{x}')= \langle \bm{q}(\bm{x}) \bm{q}^{\rm H}(\bm{x}') \rangle_{\rm E}$ is the two-point correlation tensor. Since $\boldsymbol{\mathsfbi{R}}$ is Hermitian positive-definite, its eigenvalues $\lambda^{(i)}$ are real positive that each represents a fraction of the fluctuation energy, and the eigenvectors $\{\boldsymbol{\psi}^{(i)}\}_{i=1}^{M}$ form an orthogonal basis under the inner product \eqref{eqn:inner}. 

% For a time-homogeneous statistically stationary process $\bm{q}(\bm{x},t)$, and i
In the framework of spectral POD (SPOD), the eigenvalue problem \eqref{eqn:evp} is cast as \begin{equation}
\mathcal{R} \boldsymbol{\psi}^{(i)}(\bm{x},t) = \int_{\bm{\Omega}} \int_{-\infty}^{\infty} \boldsymbol{\mathsfbi{R}}(\bm{x},\bm{x}',t,t') \boldsymbol{\mathsfbi{W}}(\bm{x}') \boldsymbol{\psi}^{(i)}(\bm{x}',t')  \, {\rm d} \bm{x}'{\rm d} t' = \lambda^{(i)} \boldsymbol{\psi}^{(i)}(\bm{x},t), 
\label{eqn:evp-spod}
\end{equation} and $\boldsymbol{\mathsfbi{R}}(\bm{x},\bm{x}',t,t') = \langle \bm{q}(\bm{x},t) \bm{q}^{\rm H}(\bm{x}',t') \rangle_{\rm E} $ is the two-point, two-time correlation tensor. With time-homogeneity, it reduces to $\boldsymbol{\mathsfbi{R}}(\bm{x},\bm{x}',\tau)$ as a function of $\tau=t-t'$, and is the Fourier transform pair of the spectral density tensor $\boldsymbol{\mathsfbi{S}}(\bm{x},\bm{x}',f) = \langle \hat{\bm{q}}(\bm{x},f) \hat{\bm{q}}^{\rm H}(\bm{x}',f) \rangle_{\rm E}$: \begin{equation}
  \boldsymbol{\mathsfbi{R}}(\bm{x},\bm{x}',\tau) = \int_{-\infty}^{\infty} \boldsymbol{\mathsfbi{S}}(\bm{x},\bm{x}',f) e^{i 2 \pi f \tau}\, {\rm d} f. 
\end{equation} Hence, $\boldsymbol{\phi}^{(i)}(\bm{x},f) = \boldsymbol{\psi}^{(i)} (\bm{x},t) e^{-i 2 \pi f \tau}$ will be the corresponding eigenmodes of the following eigenvalue problem \begin{equation}
  \mathcal{S} \boldsymbol{\phi}^{(i)}(\bm{x},f) = \int_{\bm{\Omega}}  \boldsymbol{\mathsfbi{S}}(\bm{x},\bm{x}',f) \boldsymbol{\mathsfbi{W}}(\bm{x}') \boldsymbol{\phi}^{(i)}(\bm{x}',f)  \, {\rm d} \bm{x}' = \lambda^{(i)}(f) \boldsymbol{\phi}^{(i)}(\bm{x},f), 
  \label{eqn:evp-spectral}
\end{equation} which will be solved separately for each frequency. Here $\hat{\bm{q}}(\bm{x},f)$ denotes the Fourier mode of ${\bm{q}}(\bm{x},t)$ at frequency $f$, and can be represented by the eigenfunctions $\boldsymbol{\phi}^{(i)}(\bm{x},f)$ as \begin{equation}
  \hat{\bm{q}}(\bm{x},f) = \sum_{i=1}^{\infty} \sqrt{\lambda^{(i)}(f)} \boldsymbol{\phi}^{(i)}(\bm{x},f).  
\end{equation} {In the occasion of spectral POD, the weighted inner-product in \eqref{eqn:inner}-\eqref{eqn:ops} will be a space-time integral \begin{equation}
  (\bm{q}_1, \bm{q}_2)_{\boldsymbol{\mathsfbi{W}}} = \int_{\bm{\Omega}} \int_{-\infty}^{\infty} \bm{q}_2^{\rm H}(\bm{x},t)  \boldsymbol{\mathsfbi{W}}(\bm{x}) \bm{q}_1(\bm{x},t) \, {\rm d} \bm{x} {\rm d} t. 
  \label{eqn:inner-spod}
\end{equation}}

We note that at the same frequency $f$, different eigenvectors $\boldsymbol{\phi}^{(i)}(\bm{x},f), \boldsymbol{\phi}^{(j)}(\bm{x},f)$ are orthogonal under the spatial inner product \eqref{eqn:inner} due to the symmetric positive-definiteness of $\boldsymbol{\mathsfbi{S}}(\bm{x},\bm{x}',f)$. But eigenvectors $\boldsymbol{\phi}^{(i)}(\bm{x},f_1), \boldsymbol{\phi}^{(i)}(\bm{x},f_2)$ at the same rank ($i$) associated with different frequencies are not necessarily orthogonal under space-only inner product. 

Our numerical implementation is similar to those in \cite{towne2018spectral,schmidt2020guide}. Data are sampled into blocks of sequenced snapshots (shown below is the $l$-th block)\begin{equation}
  \boldsymbol{\mathsfbi{Q}}^{(l)}= [\boldsymbol{\mathsfbi{q}}_1^{(l)}, \boldsymbol{\mathsfbi{q}}_2^{(l)}, ..., \boldsymbol{\mathsfbi{q}}_{N_{\rm FFT}}^{(l)}] \in \mathbb{R}^{N_d \times N_{\rm FFT}}, 
\end{equation} where each column $\boldsymbol{\mathsfbi{q}}_i^{(l)}$ is one snapshot.  The total number of snapshots in one block (ensemble) is $N_{\rm FFT}$. The degree of freedom of one snapshot is $N_d = N_x \times N_y \times N_z \times N_{\rm var}$, where $N_{\rm var}$ is the dimension of the vector $\bm{q}(\bm{x},t)$. In this study, we apply SPOD on the vertical component of vorticity ($\omega_z$, hence $N_{\rm var}=1$) in horizontal and vertical two-dimensional cross-sections of the flow (with either $N_z=1$ or $N_y=1$, respectively).

A discrete Fourier transform (DFT) is performed on each block $\boldsymbol{\mathsfbi{Q}}^{(l)}$ to yield \begin{equation}
  \hat{\boldsymbol{\mathsfbi{Q}}}^{(l)} = [\hat{\boldsymbol{\mathsfbi{q}}}_1^{(l)}, \hat{\boldsymbol{\mathsfbi{q}}}_2^{(l)}, ..., \hat{\boldsymbol{\mathsfbi{q}}}_{N_{\rm FFT}}^{(l)}] \in \mathbb{C}^{N_d \times N_{\rm FFT}} .
  \label{eqn:fft}
\end{equation} Then the Fourier modes are sorted according to frequency (labeled as the $k$-th discrete frequency) to form \begin{equation}
  \hat{\boldsymbol{\mathsfbi{Q}}}_{k} = [\hat{\boldsymbol{\mathsfbi{q}}}_{k}^{(1)}, \hat{\boldsymbol{\mathsfbi{q}}}_{k}^{(2)},..., \hat{\boldsymbol{\mathsfbi{q}}}_{k}^{(N_{\rm blk})}] \in \mathbb{C}^{N_d \times N_{\rm blk}}  . 
\end{equation}

The sampled spectral density at the $k$-th frequency is then $\boldsymbol{\mathsfbi{S}}_k = \hat{\boldsymbol{\mathsfbi{Q}}}_{k} \hat{\boldsymbol{\mathsfbi{Q}}}_{k}^{\rm H} /(N_{\rm blk}-1)$ and the discrete form of the eigenvalue problem \eqref{eqn:evp-spectral} is \begin{equation}
  \boldsymbol{\mathsfbi{S}}_k \boldsymbol{\mathsfbi{W}} \mathnormal{\boldsymbol{ \Phi}}_k = \mathnormal{\boldsymbol{\Phi}}_k \mathnormal{\boldsymbol{\Lambda}}_k, 
  \label{eqn:evp-dis}
\end{equation} with the weight matrix $\boldsymbol{\mathsfbi{W}}$ containing the  weights of numerical quadrature at each grid point. In practice, \eqref{eqn:evp-dis} is typically solved with the method of snapshots of \cite{sirovich1987turbulence} by replacing  $\boldsymbol{ \Phi}_k = \hat{\boldsymbol{\mathsfbi{Q}}}_{k} \boldsymbol{\Psi}_k $ such that \eqref{eqn:evp-dis} becomes an equivalent eigenvalue problem \begin{equation}
  \frac{1}{N_{\rm blk}-1} \hat{\boldsymbol{\mathsfbi{Q}}}_{k}^{\rm H} \boldsymbol{\mathsfbi{W}} \hat{\boldsymbol{\mathsfbi{Q}}}_{k} {\boldsymbol \Psi}_k = {\boldsymbol \Psi}_k \boldsymbol{\Lambda}_k. 
  \label{eqn:evp-dis-snap}
\end{equation} that has a much smaller dimension  {when $N_{\rm blk} \ll N_d$ is true.} Hence, the eigenmodes of $\boldsymbol{\mathsfbi{S}}_k$  {are recovered as} $\tilde{\boldsymbol{ \Phi}}_k = \hat{\boldsymbol{\mathsfbi{Q}}}_{k} \boldsymbol{\Psi}_k  \boldsymbol{\Lambda}_k^{-1/2}$ such that the eigenvalue decomposition is \begin{equation} 
  \boldsymbol{\mathsfbi{S}}_k =\tilde{\boldsymbol{ \Phi}}_k \boldsymbol{\Lambda}_k \tilde{\boldsymbol{ \Phi}}_k^{\rm H} = \sum_{i=1}^{N_{\rm blk}} \lambda_{k}^{(i)} \tilde{\boldsymbol{\phi}}_k^{(i)} (\tilde{\boldsymbol{\phi}}_k^{(i)})^{\rm H}. 
\end{equation} The physical meaning of the spatial modes $\tilde{\boldsymbol{ \Phi}}_k(\bm{x})$ can be interpreted as either the eigenvector of the spectral density tensor $\boldsymbol{\mathsfbi{S}}_k$ or the left singular vector of the Fourier mode $\hat{\boldsymbol{\mathsfbi{q}}}_{k}$, at a discrete frequency $f_k$.

SPOD takes advantage of extracting spatial modes that evolve at a single frequency from  a time-resolved database as in table \ref{table1}. It was applied to analyse stratified wakes by \cite{nidhan2020spectral,nidhan2022analysis}, who showed that it can successfully extract the large-scale VS motions and the associated characteristic frequency. In oceanography applications, \cite{zeiden2021broadband} applied a similar approach called empirical orthogonal functions (EOF) therein to the flow past an island and successfully separated the vortical modes and the tidal modes. But in their case, the EOFs are fit to three mooring points instead of the bulk of the flow, hence are different from ours where the eigenmodes will be emphasised as global modes. 

When converting the time series into Fourier modes in \eqref{eqn:fft}, Welch's method \citep{welch1967use} is used to reduce the variance of the spectrum, with $N_{\rm FFT} = 512$ snapshots in each block, and a Hamming window on each block to enforce periodicity. An overlap ratio of $50\%$ ($N_{\rm ovlp}=256$) between two sequential blocks is chosen to offset the effect of low weights near the edges of the window. We end up with 13 blocks and the ensemble average will be taken over all blocks to obtain SPOD eigenspectra. The convergence of the method is checked via reducing the frequency resolution to $N_{\rm FFT} = 256$, or reducing the total number of snapshots from $N_t=4000$ to 3000, and 2000. A high confidence level is obtained for the largest six eigenvalues as well as the sum of all eigenvalues, at each frequency. This follows the fact that for general eigenvalue-revealing algorithms, large eigenvalues converge faster. And it is noted that, in the present wakes, converging high-rank SPOD eigenvalues  {with much smaller magnitudes} is still challenging even with $N_t=4000$ snapshots. In this paper, no particular analysis will be conducted for higher than the sixth eigenvalue at any frequency. 

% \jy{jy: I put these comments because I saw people discussing in papers (S)POD modes ranked 10 or even 100, without justifying their convergence. The convergence of high-rank EVs seem to be an issue whether it is spectral POD or not. POD EV's are easier though. For SPOD of jets (OTS et al.), I remember the use of $N_t=10000$ barely converged the third or the second EV at each frequency, since their jets are more turbulent.}

In this database, a constant maximal Courant–Friedrichs–Lewy (CFL) number is kept during the simulation, which results in an uneven (but almost uniform) time spacing of snapshots. To obtain uniformly spaced data for DFT, a piecewise cubic Hermite interpolation (PCHIP) is performed in time. 

% \jy{In the SPOD analysis, the vertical vorticity on different horizontal planes ($\omega_z(x,y,t;z)$) is chosen as the quantity of interest, base on the observation in figures \ref{fig:3dvis}-\ref{fig:omgz_mean} that the wakes exert predominantly Q2D vortical motion, the shedding and evolution processes of which is well represented by $\omega_z$. Moreover, the focus of this work is on the influence of increasing rotation strength on the horizontal motion, which indeed gradually constrains the flow to be around the vertical axis at the large scales and significantly enhance $\omega_z$ as will be shown later. In the rotating cases Bu25 and Bu1, the stability of the anticyclones (vortices that rotate in direction opposite to frame rotation) will also be studied (section \ref{cyclones}) with $\omega_z$ being one of the most important quantity of determination, hence we apply $\omega_z$ rather than other vortex identification criteria for overall consistency. }

%%%%%%%%%%%%%%%%%%%%%%%%%%%%%%%%%%%%%%%%%%%%%%%%%%%%%%%%%%%%%%%%%%%%%%
\subsection{SPOD eigenspectra and vortex shedding frequencies} \label{eigenspectra}

% \jy{stress the global frequency. stability: how the flow looks around them}

\begin{figure}[t!]
  \captionsetup[subfloat]{farskip=-12pt,captionskip=1pt}
  \subfloat[]{{\includegraphics[width=.45\linewidth]{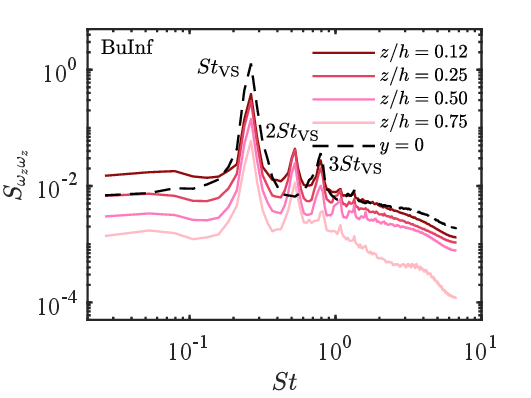}}}% 
  \subfloat[]{{\includegraphics[width=.45\linewidth]{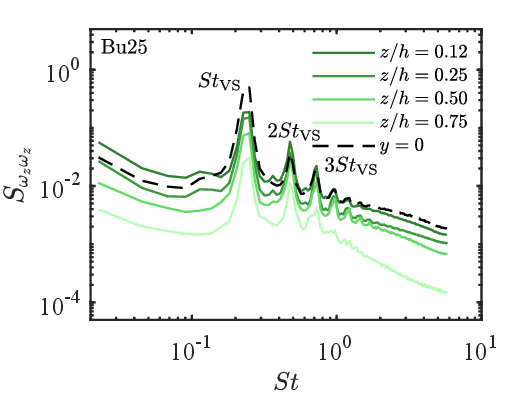}}}% 
   \\
  \subfloat[]{{\includegraphics[width=.45\linewidth]{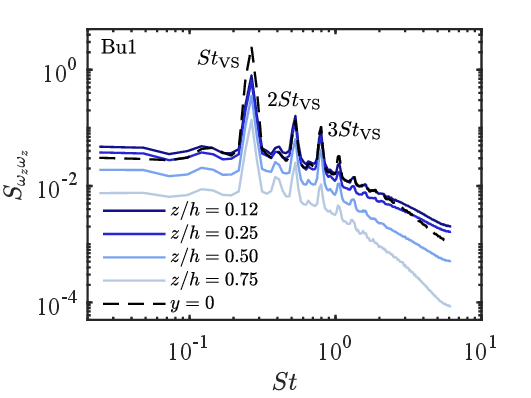}}}% 
  \subfloat[]{{\includegraphics[width=.45\linewidth]{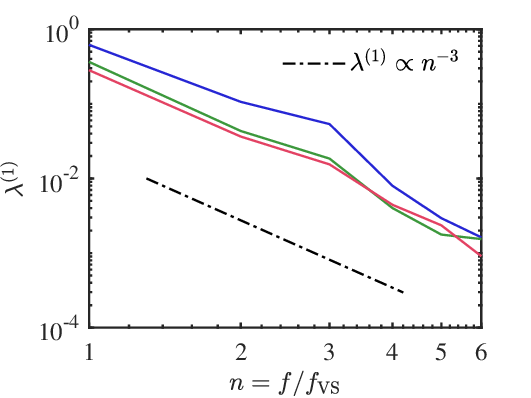}}}% 
  \caption{(a-c) Global power spectra $S_{\omega_z \omega_z}$ of $\b.mega_z$ as a function of Strouhal number ($St=fD/U_{\infty}$). Spectra are shown at four horizontal planes $z/h=0.12,0.25,0.50,0.75$, and the central vertical plane at $y=0$. The VS Strouhal number is marked as $St_{\rm VS}$ as well as the VS harmonics. The values take $St_{\rm VS}=0.264, 0.249,$ and $0.266$ in (a-c),  {for cases BuInf, Bu25, and Bu1,} respectively. (d) Decay of the largest eigenvalue at the VS frequency $f_{\rm VS}$ and the harmonics (indexed by $n=f/f_{\rm VS}$) for the horizontal plane $z/h=0.25$ (results are similar for other locations). Color codes same as in (a-c).} 
  \label{fig:eigens}
\end{figure}

Figure \ref{fig:eigens}(a-c) shows the global vertical enstrophy spectra $S_{\omega_z \omega_z}(f)$ at different 2D planes for BuInf, Bu25, and Bu1, respectively. The spectral density at a discrete frequency $f_k$ is computed by summing over all SPOD eigenvalues at this frequency: \begin{equation}
  S_{\omega_z \omega_z}(f_k) = \sum_i^{N_{\rm blk}} \lambda^{(i)}(f_k) = {\rm tr}(\boldsymbol{\mathsfbi{S}}_k \boldsymbol{\mathsfbi{W}} ) = \int_{\bm{\Omega}} \boldsymbol{\mathsfbi{S}}(\bm{x},\bm{x},f_k) \, {\rm d}\bm{x},  
\end{equation} and is the spectral density of the area-integrated squared $\omega_z$. It is independent of SPOD.

For all three cases and all planes examined, the spectra display strong harmonic spikes, with the strongest peak defined as the VS frequency ($f_{\rm VS}$) in each plane. The VS frequency is independent of the vertical location of the four horizontal planes, and is also the same in the central vertical plane, {even though performing SPOD on separate planes allows the freedom of selecting different frequencies}. This indicates that for each case, $f_{\rm VS}$ is a {\rm global constant} (for the heights examined), {and the VS modes are three-dimensional \rm{global modes}}. The VS Strouhal number is $St_{\rm VS} = f_{\rm VS} D/U_{\infty} = 0.264,0.249,0.266$ for BuInf, Bu25, and Bu1, respectively. 
It is noted that, since the characteristic frequency of the global mode ($f_{\rm VS}$) does not depend on the height or the local hill diameter, normalising it with a different length scale than the hill base diameter $D$ would just make $St_{\rm VS}$ different by a scalar multiple. However, as will be discussed later, the numerical values of $St_{\rm VS}$ using $D$ correspond well to that of vortex shedding from a circular cylinder. 
Also, since $D$ determines the size of the largest scales in the flow, it is a natural choice for the normalizing scale. 
In the eigenspectra, the successive peaks are harmonics of the VS frequency at $2St_{\rm VS}, 3St_{\rm VS},$ and so on. 

\cite{perfect2018vortex} found that whether the VS frequency is a global constant or is controlled by the local hill diameter, depends on a non-dimensional parameter: Burger number ($Bu$). The Burger number characterises  the relative importance of two counteracting mechanisms for vertical coupling: rotation and stratification. When $Bu$ is small (rotation is strong), the bulk of the flow adjusts to be around the axis of rotation quickly, and geostrophic balance is established, where the vertical variation is minimized. As $Bu$ is increased, the vertical intercommunication is progressively weakened by stratification.
{\cite{perfect2018vortex} use the diameter at half-height to be the characteristic horizontal scale so that their values of Rossby number ($Ro^*$) and Burger number ($Bu^*$) are related to our values by $Ro^* = 2 Ro$ and $Bu^* = 4 Bu$.
They performed a number of simulations and found that ${Bu}^{*}_{\rm cri}=5.5$, equivalently ${Bu}_{\rm cri} = 1.4$, is a regime-separation criterion below which the rotation is strong enough to couple different layers to form vertically aligned vortices. Also,  when  ${Bu^*} > 12$, equivalently ${Bu}  > 3$,  they found stratification to be  more prominent so that vortices are shed at different layers relatively independently. Turning to the present results, in each of the  cases at $Fr=0.15$  whose $Ro$ span unity to infinity, the VS frequency is independent of height,  indicating a potential disagreement between \cite{perfect2018vortex}  and our results.}
%\jy
{On the other hand, for Froude numbers similar to $Fr=0.15$ in \cite{perfect2018vortex},  almost all the cases (see figure 4, therein)  were labeled as `vertically coupled shedding' and had strong rotation with $Ro$ between 0.025 and 0.25. Only their weakest rotation case with $Ro = 0.25$ was labeled as `vertically decoupled shedding'. 
% In light of the present results, it is possible that  $Fr$ smaller than 0.15 is required for stratification to be sufficiently strong for vertically decoupled shedding. 
 It is also worth noting that, in the ROMS simulations, vertical motions and pressure correlations are quite approximate (specially in the near wake where vortex shedding is accompanied by small-scale turbulence)   since the momentum equation in that direction is reduced to a hydrostatic balance, and the pressure field might play an important role in coupling VS.
}

%didn't have cases equivalent to or higher than $Ro>1$ ($Ro$ after converted to our definition; see figure 4 therein), hence all cases there at stratification level around $Fr=0.15$ are labeled as `coupled' shedding, which is not in contradiction to our findings. It is possible that at this relatively high $Fr=0.15$, if $Ro$ is increased, VS will stay coupled. The second possibility is that, in the ROMS simulations, vertical motions and pressure correlations are not well-resolved since the momentum equation in that direction is reduced to a hydrostatic balance, and the pressure field might play an important role in coupling the VS. }
% Also, at higher $Fr$, the hydrostatic assumption is less accurate than at very low $Fr$ based on order of magnitude arguments (see appendix). 

Nevertheless, the fact that the modes extracted by SPOD are global modes, and they evolve at the same frequency, implies that the large-scale vortices at $Fr = 0.15$ are horizontally and vertically coherent, as opposed to the asymptotic limit of vertically uncoupled layered dynamics in strongly stratified flows. Our findings indicate that the stratification of $Fr=0.15$ is not strong enough to vertically decouple the vortex dynamics in the wake of the conical seamount, regardless of the presence of rotation, and inclusion of $Fr$ is necessary in addition to $Bu$. The pure Burger number criterion has the limitation that, for instance, with no rotation and small stratification, $Bu$ will be far larger than $Bu_{\rm cri}$ but the VS frequency can still be a global constant. The $Fr$ that  marks the   transition from vertically coupled to decoupled VS  in non-rotating hill wakes is subject to future research.

\cite{boyer1987stratified} studied experimentally the wake behind a conical obstacle  in linearly stratified rotating flows. Their parameters spanned $0.08<Fr<0.28$, $0.06<Ro<0.4$, and for three Reynolds numbers $Re_D=380,760,1140$. Based on the measurement on single horizontal cross-sections, they found the VS Strouhal number to be only a weak function of both $Re$ and $Ro$. The robustness of the VS frequency to rotation strength is also observed numerically in this work, in a similar $Fr$-$Ro$ regime but at a turbulent Reynolds number $Re_D=10\,000$. Even though their Strouhal numbers are measured as the vortex advection velocity divided by the mean separation of two same sign vortices, and ranged from $0.20<St<0.35$, our VS Strouhal numbers still present a good quantitative agreement with theirs. Moreover, we interpret the VS frequency revealed by SPOD as the characteristic frequency of the most energetic global mode, which also agrees with single-point frequency measurements in the intermediate wake ($x/D>3$, not shown). 

The values of $St_{\rm VS}$ for all three cases are close to $St=0.2665$, which is the $Re \rightarrow \infty$ asymptote of the $St$-$Re$ relationship in low Reynolds number 2D cylinder wakes proposed by \citep{williamson1998series}. Note that their relation $St= 0.2665-1.018\sqrt{Re}$ is given for $50<Re_D<180$ which is before the transition to 3D wakes. This transition happens at around $Re=188.5$ according to a global Floquet instability of the periodic wake \citep{barkley1996three}. As a result, the $St$-$Re$ relationship experiences a discontinuity as a sudden jump of $St$ during this transition \citep{williamson1998series,fey1998new}, and the asymptote of $St=0.2665$ {is not} reached in three-dimensional cylinder wakes. \cite{fey1998new} showed that the maximum VS Strouhal number in a cylinder wake of about $St=0.21$ is reached right before the onset of the Kelvin-Helmholtz instability in the shear layer. We interpret the observed values of VS frequencies in our hill wakes as a saturation of the Q2D VS, which {will not} be observed in three-dimensional cylinder wakes at this Reynolds number and above. This interpretation is consistent with the finding in \cite{boyer1987stratified} that the robustness of $f_{\rm VS}$ to rotation rate is not affected by the Reynolds number, in their low-to-moderate Reynolds number experiments.

\begin{figure}[t!]
  \includegraphics[width=0.975\textwidth]{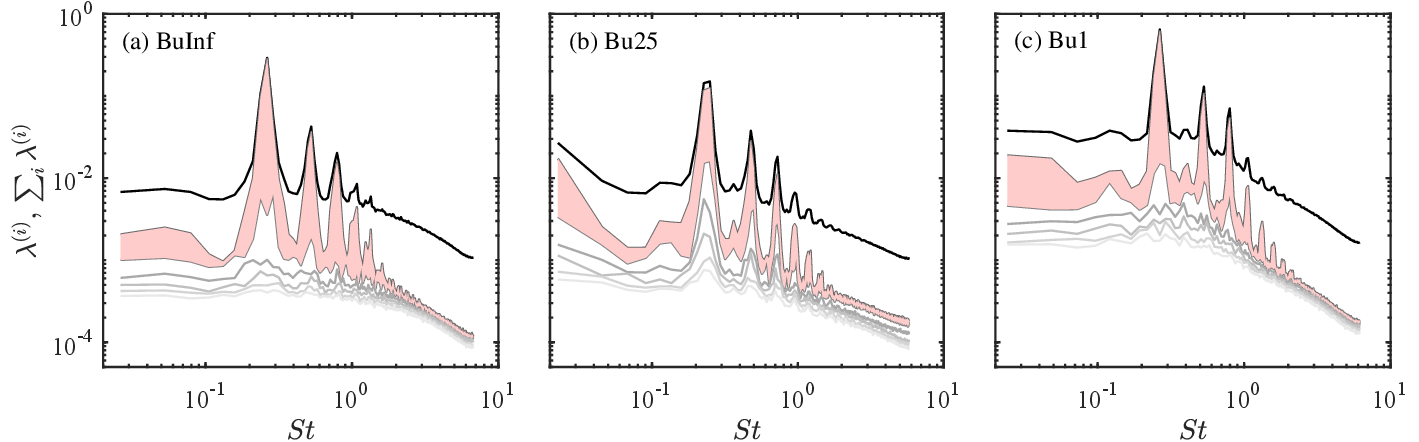}
  \caption{SPOD eigenspectra: (a) BuInf, (b) Bu25, and (c) Bu1. Horizontal plane at $z/h=0.25$ is shown (results are similar for other locations). From top to bottom are the summation of all eigenvalues (darkest, spectral density as in figure \ref{fig:eigens}(a-c)), and the first to the sixth eigenvalues (from dark to light: $i=1,2,3,4,5,6$), as a function of Strouhal number ($St=fD/U_{\infty}$). The difference between the first and the second eigenvalue is filled with color.} 
  \label{fig:eigengap}
\end{figure} 

The enstrophy distribution among different eigenvalues is an important measure of the complexity of the system. Figure \ref{fig:eigengap} shows the sum of all eigenvalues, and also the first to the sixth eigenvalues (from dark to light) according to their absolute value. For frequencies at or close to the VS frequency and its harmonics, the first eigenvalue  accounts for most of the enstrophy, and is an order of magnitude larger than the second eigenvalue. However, for larger frequencies ($St>2$), the dominance of the leading eigenvalues is lost, as the degree of freedom for small-scale motions is increased. The significance of low-rankness in the spectra is therefore two-fold. There are strong harmonic spikes in the SPOD eigenspectra in figure \ref{fig:eigens}(a-c), implying that a great portion of enstrophy is contained in the large-scale VS motions. For the VS shedding frequency and its harmonics, the first two eigenvalues contribute the most of the enstrophy. 

Moreover, the enstrophy distribution among the harmonics is shown in figure \ref{fig:eigens}(d). The decay of the first eigenvalue $\lambda^{(1)}$ as a function of the integer harmonic index $n=f/f_{\rm VS}$ is plotted. Similar to figure \ref{fig:eigengap}, one horizontal plane $z/h=0.25$ is chosen, but the results are qualitatively similar for the other three selected horizontal planes. For all three Burger numbers,  {$\lambda^{(1)}$ decays approximately as a power progression as $n^{-3}$, 
% with an index (-3) close to that in the decay of distinct POD eigenvalues in post-transition 3D cylinder wake at $Re_D=185$ \citep{ma2002low}, 
which is slower than the geometric decay of distinct POD eigenvalues  
% and Fourier coefficients of harmonic perturbations \citep{duvsek1994numerical}, both 
in 2D cylinder wakes \citep{noack2003hierarchy}.} Nevertheless, higher harmonics contain a progressively smaller amount of enstrophy. Additionally, by comparing the absolute value of $\lambda^{(1)}$ in figure \ref{fig:eigens}(d) and $\sum_i \lambda^{(i)}$ in {figure} 
% \ref{fig:eigens}(a-c) 
\ref{fig:eigengap}, it is generally true that vertical enstrophy is increased as the rotating rate increases. This is an effect of  system rotation that promotes vortical motion around the rotation axis as will be discussed in more detail in section \ref{cyclones}. 

In all, the large scales of the vortical motion can be well represented by a few characteristic frequencies and the associated leading modes. Such low-rank behaviour suggests the possibility of reduced-order modelling of stratified hill wakes even at large Reynolds numbers.

\subsection{SPOD eigenfunctions and large-scale global modes} \label{eigenmodes}

\begin{figure}[p]
  \captionsetup[subfloat]{farskip=-6pt,captionskip=1pt}
  \subfloat[]{{\includegraphics[width=.45\linewidth]{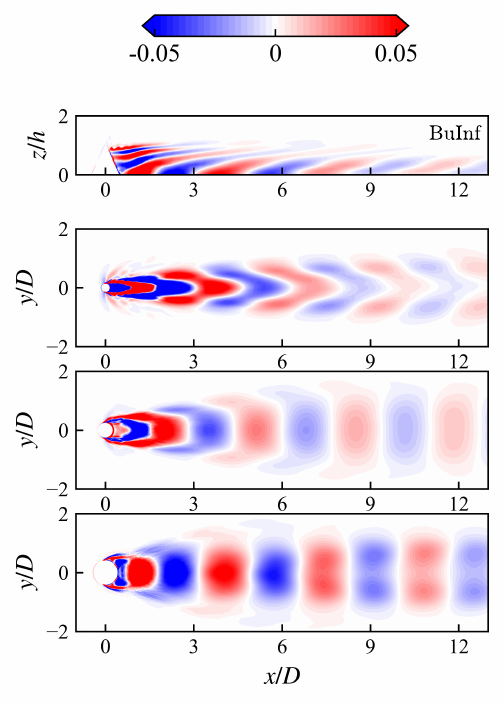}}}% 
  \subfloat[]{{\includegraphics[width=.45\linewidth]{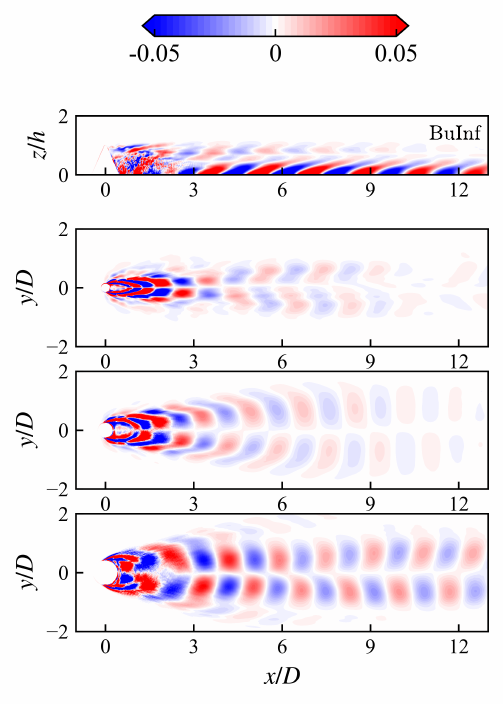}}}% 
   \\ 
  \subfloat[]{{\includegraphics[width=.45\linewidth]{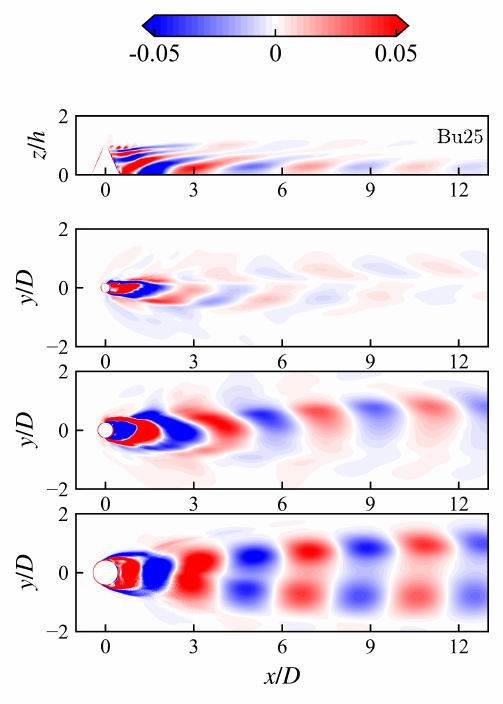}}}% 
  \subfloat[]{{\includegraphics[width=.45\linewidth]{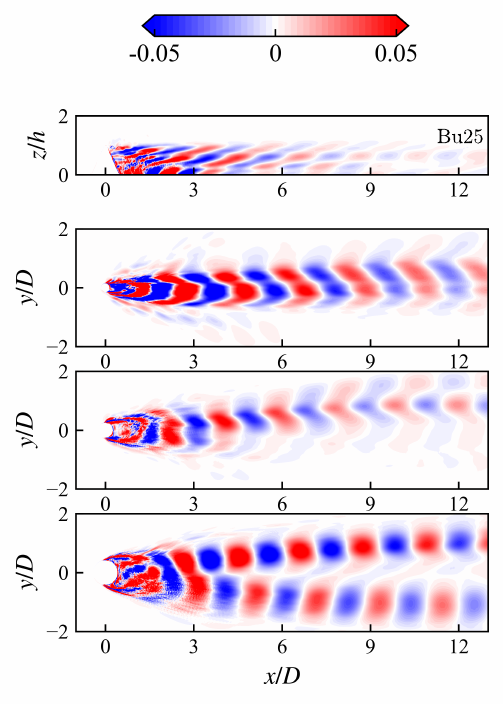}}}% 
  \caption{Leading SPOD eigenmodes corresponding to $f_{\rm VS}$ (a,c) and $2f_{\rm VS}$ (b,d), for cases BuInf (a,b) and Bu25 (c,d). The plotted quantity is the real part of SPOD eigenfunction on each plane, normalized by the maximum value in the plane. In each figure, the top plane is the vertical plane, $y=0$, and the lower planes are, from the second row to bottom, horizontal planes at $z/h=0.75, 0.50, 0.12$. } 
  \label{fig:spod}
\end{figure} 

\begin{figure}[tp]
  \subfloat[]{{\includegraphics[width=.45\linewidth]{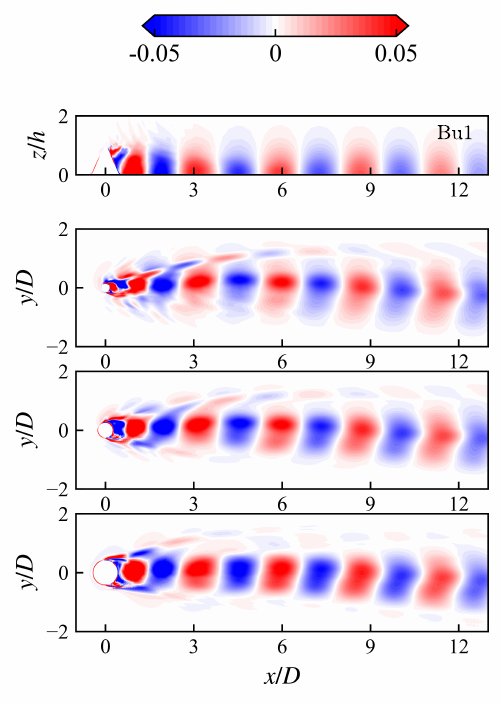}}}% 
  \subfloat[]{{\includegraphics[width=.45\linewidth]{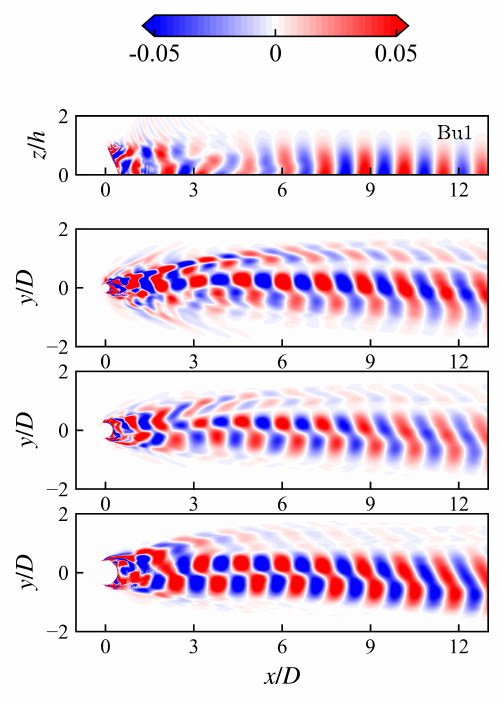}}}% 
  \caption{For case Bu1. Caption same as in figure \ref{fig:spod}.}
  \label{fig:spoe}
\end{figure}
 
Apart from the temporal characteristics uncovered by SPOD eigenspectra, the SPOD eigenmodes represent energetic flow structures that are coherent in space and time and, therefore, dynamically important. {The dominant modes in each case are found to be the vortex shedding modes, which are three-dimensionally coherent and characterised by the VS frequency ($f_{\rm VS}$) and its superhamonics.} 

The real parts of the leading SPOD modes (corresponding to the largest eigenvalue at each frequency) are plotted in figures \ref{fig:spod}-\ref{fig:spoe}, for two frequencies $f_{\rm VS}$ and $2f_{\rm VS}$ and on various 2D planes. The eigenfunctions are free up to a scalar multiple and are normalised by the largest modulus of the spatial mode in the same plane. The interpretation of the magnitude of the eigenmodes (or the darkness of the colors) as the intensity of the structures is only meaningful when the comparison is made within the same plane.  

For case BuInf (no rotation), the VS mode in figure \ref{fig:spod}(a) is reminiscent of the marginally stable global modes found in the linear stability analysis of low Reynolds number cylinder wakes \citep{barkley1996three,mittal2008global}. However, the phases of the global mode are different in different planes, with the phases of higher planes leading. This is clearer in the vertical plane ($y=0$), which shows the tilting and elongation of the structures as an effect of stratification. Note that the vertical coordinate is normalised by the hill height $h$ which is \st{about} 0.3 times the base diameter; in un-normalized coordinates, the angle of the slanted {`tongues'} is very shallow. The tilting angle from the horizontal of the structures in the top panel of figure \ref{fig:spod}(a) ranges from about $8^{\circ}$ close to the bottom wall to $2^{\circ}$ near the top, with an average of about $4^{\circ}$ which is almost constant during the evolution. Although the VS structures are not vertical, they still evolve cohesively at $f_{\rm VS}$ and experience little change  during the advection. 
%{The tilting angle of the structures from horizontal is between $xx\deg $ and $xx \deg$, which is also preserved during streamwise advection. }
In the later section \ref{tracking}, the lack of change of the tilting angle will be shown to be a result of a roughly constant advection velocity of the vortices. Similar slanted mode as in figure \ref{fig:spod}(a) was found in a stratified sphere wake by \cite{chongsiripinyo2017vortex} (referred as `surfboards' therein), but whether it is a common feature of stratified bluff-body wakes is subject to future research. 

Moreover, the standing-wave-like spatial modes of $f_{\rm VS}$ in figure \ref{fig:spod}(a) are symmetric about $y=0$, representing the advection of the perturbation vorticity $\omega'_z=\omega_z - \langle \omega_z \rangle $. The projection of $\omega'_z$ on the $f_{\rm VS}$ mode has a period of $T_{\rm VS}$ and the streamwise wavelength of the mode is interpreted as the average spacing of the K{\'a}rm{\'a}n vortices. The appearance of the symmetric $f_{\rm VS}$ mode on the antisymmetric mean flow (see figure \ref{fig:omgz_mean}d),  in analogy to low Reynolds number 2D cylinder wakes \citep{kumar2006effect}, can be viewed as an accompaniment to the symmetry-breaking bifurcation from a steady wake to periodic shedding.

The modes of $2 f_{\rm VS}$ in figure \ref{fig:spod}(b) are antisymmetric, with a wavelength  about half of that in the $f_{\rm VS}$ mode. A result of this reflection antisymmetry is that the eigenspectrum at the vertical plane (black dashed line in figure \ref{fig:eigens}{(a)} {does not} exhibit a peak at $2f_{\rm VS}$, unlike the spectra at horizontal planes. The antisymmetry implies a zero magnitude of the $2f_{\rm VS}$ mode  at the centreline ($y=0)$, hence the top row in {figure} \ref{fig:eigens}(b) does not imply any dynamical importance. 

% \jy{(jy: previous statements on the degeneracy are not good; I've changed to the below; comparison to POD eigenvalue degeneracy is removed since we discussed it in the Madrid proceeding.)} 

 {The imaginary parts of the eigenmodes are not shown, which are different from the real parts by a streamwise phase shift of $\pi/2$, and are therefore representing the same VS dynamics. The phase shift {between two} standing-wave-like eigenmodes with the same streamwise wavelength (such as the real and imaginary parts of the $f_{\rm VS}$ mode here) {is essential for them to accommodate one} traveling-wave-like structure (such as the advecting VS motion) as noted by \cite{rempfer1994evolution}. 
% The symmetries of the modes shown in figure \ref{fig:spod}(a,b) are linked to the spatial-temporal symmetry that the non-rotating wake (BuInf) possesses, as in a periodic K{\'a}rm{\'a}n wake. The spatial-temporal symmetry for $\omega_z$ reads (at a constant $z/h$ location) \begin{equation}
  % {\omega}_{z}(x,y,t+T_{\rm VS}) = {\omega}_{z}(x,y,t); \; {\omega}_{z}(x,y,t+T_{\rm VS}/2) = -{\omega}_{z}(x,-y,t), 
% \end{equation} and each correspond to the reflectional symmetry of the $f_{\rm VS}$ mode, and the reflectional antisymmetry of the $2f_{\rm VS}$ mode, respectively. 
}

For cases Bu25 and Bu1, the reflectional symmetry is broken by the Coriolis force and, unlike BuInf,  the peak at $2f_{\rm VS}$ is observed in the eigenspectra at the vertical plane (black dashed lines). The rightward ($-y$) shift of the vortex wake is consistent with the direction of the Ekman veering near the bottom by the unbalanced mean pressure gradient. This asymmetry can also be found in the SPOD modes in figures \ref{fig:spod}(c,d) and \ref{fig:spoe}(a,b), where the cyclones (on the left side of the bulk of the wake, looking from above) are preferred and amplified, as compared to the anticyclones. However, in the vertical mode (first row in figure \ref{fig:spod}(c,d)), slanted structures are still statistically significant, even though figure \ref{fig:3dvis}(b) shows that, for case Bu25, some anticyclones have already become columnar. This suggests that the stratification in Bu25 is still dominating, and rotation has not been able to modify the spatial organisation of the structures. However, the presence of rotation significantly alters the structure inside the anticyclones, which will be analysed in section \ref{cyclones}.

For case Bu1 (strong rotation), the global modes in figure \ref{fig:spoe}(a-b) show excellent vertical alignment, agreeing with the shed `columns' in figure \ref{fig:3dvis}(c).  In the vertical plane (see the first row in figure \ref{fig:spoe}a), both cyclonic and anticyclonic structures extend slightly higher than the obstacle peak, whereas the structures in cases Bu25 and BuInf are limited to below the obstacle peak, signifying that the obstacle's range of influence is vertically increased by increasing rotation. We note that those `columns' are different from the typical Taylor column in rotating flow over obstacles. A Taylor column has an infinite height on the top of a finite object that generates it, whereas the columnar global mode in case Bu1 has a finite height. In figures \ref{fig:spoe}(a,b) it can be seen that the parts of the global mode  {near the hill ($0<x/D<1$)} are smaller and slanted, and they  eventually become organised into tall `columns' during their advection downstream as rotation is experienced, instead of being columnar at the generation. 

% \jy{(jy: a question I just have: why the asymmetry between AVs and CVs in terms of height and size, is not seen in the top row of SPOD modes in fig. 7a? I guess the answer is the SPOD mode at $y=0$ has a mix of and is an average of both AVs and CVs.)}

{Finally, the relation between the global VS modes found in the present stratified wakes and the global modes in 2D K{\'a}rm{\'a}n wakes is worth further exploration.   The VS modes observed in each 2D horizontal plane of the non-rotating case (BuInf, see figures \ref{fig:spod}-\ref{fig:spoe}) are visually similar to those observed in 2D K{\'a}rm{\'a}n wakes, and the associated VS Strouhal number ($St_{\rm VS}$) is close to the $Re\rightarrow \infty$ limit of the $St$-$Re$ relation of 2D cylinder wakes. It is worth pointing out that strong stratification is crucial for the similarity between the present high Reynolds number wakes and 2D K{\'a}rm{\'a}n wakes. Alternating shedding vortices are not seen at Reynolds number similar to $Re_D=10\,000$ in unstratified cylinder wakes \citep{fey1998new} and stratified sphere wakes at $Fr>O(0.5)$ \citep{pal2016regeneration}, where the attached shear layer will have lost its stability.

On the other hand, the slanted, forward-tilting vertical structures seen in the $y=0$ plane of BuInf case are suggestive of vertical coherence of horizontal vortical motion, which is absent for 2D flows. As a part of the global mode, the VS frequency and spatial mode in each horizontal plane depends rather on the hill base diameter than on the local hill diameter at the same height -- another difference from typical 2D K{\'a}rm{\'a}n wakes. }   

{The change of the titling angle of the global mode as $Ro$ decreases until the vortex structures are fully upright at $Ro=0.15$ adds an additional aspect -- the influence of rotation. Even though the vertical alignment of the structures is varied, the dominant modes are still the VS modes with the shedding frequencies being similar in each case, indicating that the VS motions at present $Fr=O(0.1)$ are relatively robust to rotation. Slanted vortical structures as in BuInf were also observed in stratified non-rotating wakes past a sphere by \cite{pal2016regeneration,chongsiripinyo2017vortex}. Whether these structures are common in strongly stratified wakes past a body of revolution with vertically varying diameter and how these structures are modified by further increase in  stratification deserves future study.}

\section{Vortex centre tracking and advection velocity} \label{tracking}

% \st{In the previous section, we take a macroscopic view of the coherent structures that are characteristic features of the flow.}
% \Sutanu{Don't overuse `we ....'.} 
The previous section on coherent structures was a macroscopic view that was enabled by the extraction of  global SPOD modes.
  In this section, we will take a microscopic (at the level of individual vortices) view of the cyclonic (negative, note the Southern Hemisphere setting) and anticyclonic (positive) vortices. To do so, individual vortex centers will be identified and, by computation of statistics conditioned to them, various properties will be diagnosed on an ensemble-average basis: 
vortex advection velocity discussed in this section and, in section~\ref{cyclones}, the  vorticity distribution inside the vortices and furthermore the stability of wake vortices as inferred by the application of linear-theory-based stability criteria of varying complexity. 
%Note that throughout this paper,  the terminology `vortex' is used  instead of `eddy' for the identified coherent structure so as to highlight its long-lasting nature contrary to  a turbulent-flow eddy that breaks up within a \jy{finite turn-over time scale} \st{so-called large-eddy time scale}.
%\jy{jy: smaller eddies also have their scale-dependent turn-over times.}
%o refer to the vortical structures. Those structures have similar vorticity distribution to idealized vortex models (see section \ref{cyclones}), and don't really totally dissipate or break up within the computational domain, even the flow is turbulent. On the other hand, eddies are more frequently referred to in other turbulent dissipative flows where the lifetime is finite compared to the length of observation.  

%Tracking the evolution of individual flow structures has always been an interest in the research of turbulent shear flows where intense vortex structures account for a large portion of enstrophy in the flow and are significant to the dynamics. Especially in spatially evolving wakes, the advection and evolution of shed vortices are essential in modelling these flows. It is also of great importance to the application of Taylor's hypothesis in turbulence measurements of the wakes, which converts the measured temporal spectra at a single point to spatial spectra.
{ The advection velocity in turbulent wakes past the near-wake stage can be taken to be close or equal to the freestream velocity $U_{\infty}$, e.g., beyond $x = 6D$ in the study by \cite{vandine2018hybrid} of non-rotating wakes at $Fr \geq O(1)$.}  Whether this constancy holds everywhere in the flow, and whether there is any asymmetric advection between the cyclonic and anticyclonic sides, requires clarification for geophysical wakes.  
%On the other hand, in the context of geophysical wakes, the stability of anticyclonic vortices that has implication for turbulence generation requires the study of individual vortices. 

\begin{figure}[tp]
  \subfloat[]{{\includegraphics[width=.45\linewidth]{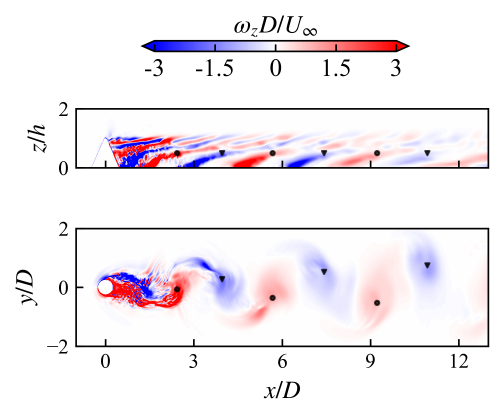}}}% 
  \subfloat[]{{\includegraphics[width=.45\linewidth]{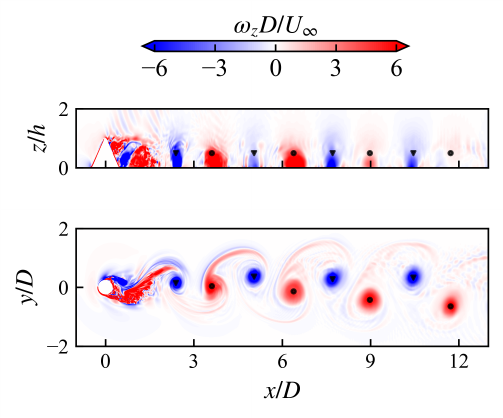}}}% 
  \caption{Visualisation of identified vortex centers, for cases (a) BuInf and (b) Bu1. Circles and triangles mark the centres of positive and negative vortices, respectively. Shown are the centres identified in a horizontal plane at $z/h=0.50$ (bottom row) and their projection in the vertical central plane at $y=0$ (top row). {The snapshots are taken at non-dimensional time instants $tU_{\infty}D= 44.9, 38.1$ counted after the fully developed state is reached, for (a) and (b), respectively.}} 
  \label{fig:vor_ctr_viz}
\end{figure}

% We take advantage of our time-resolved simulation database {and apply} a method to identify vortex centres in each snapshot and track them in time. 
{The present time-resolved database enables temporal tracking of the vortices to study their behaviour during the evolution. We apply a clustering method -- mean shift to extract the vortex centres in horizontal snapshots, and then follow each identified centre in time.} % as an individual vortex track.  
{An example of identified vortex centres is illustrated in figure \ref{fig:vor_ctr_viz}. Those centres are identified in one horizontal plane ($z/h=0.50$) which is shown in the bottom row, and their projection onto the vertical central plane ($y= 0$) is shown in the top row.} 
Symbols represent centres and are superposed on the vorticity field. To avoid the turbulent near-wake region where the wake vortices have not yet fully formed, the domain of vortex tracking starts at $x/D=2$.
%(not shown). 
The outflow region ($13<x/D<15$) is also not considered for vortex tracking, to avoid  {possible errors induced by the convective boundary condition}. The implementation of the method is presented in detail in Appendix \ref{app1}. 

\begin{figure}[tp]
  \includegraphics[width=.6\linewidth]{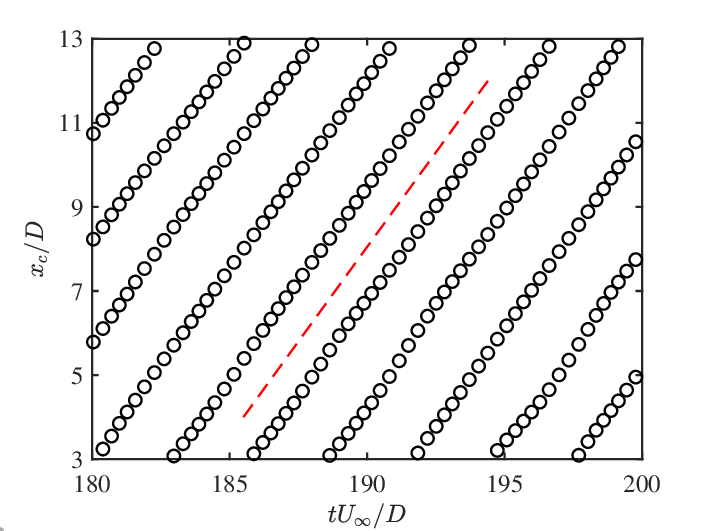}% 
  \caption{The $x_c$-$t$ diagram of the trajectories of centers of positive vortices at $z/h=0.50$ in case Bu1. Each black circle marks the instantaneous location $x_c$ of a positive vortex centre, at a time $t$. The red dashed line has a non-dimensional slope of 0.9, corresponding to a vortex advection velocity of $0.9 U_{\infty}$. For clarity, vortex centres are plotted every five snapshots.} 
  \label{fig:vor-ctr} 
\end{figure}

An example of the trajectory of vortex centres is shown in figure \ref{fig:vor-ctr} for case Bu1 at $z/h=0.50$. Each trajectory is a  sequence of streamwise locations of vortex centres $x_c$ as a function of time $t$, and the local trajectory slope gives the local vortex advection velocity. It can be seen in figure \ref{fig:vor-ctr} that the local slope is almost a constant throughout the  downstream advection and the $x_c$-$t$ trajectories are very close to straight lines. 

In order to estimate the average vortex advection velocity ($U_c$) using all available data, a linear fit of each trajectory in the $x_c$-$t$ diagram is performed to obtain the slope, and  the advection velocity at a certain height ($z/h$)  is the ensemble average over all trajectories at the same height. Since the advection velocity changes little as $x$ increases, only trajectories that last longer than 50 snapshots (around one VS period) are considered to exclude the momentary tracking of  small vortices other than the K{\'a}rm{\'a}n vortices. No significant difference is found in the results by changing this number to 25. In total, more than 80 trajectories are extracted for either positive or negative vortices in each case.
%which roughly agrees with the number of vortex shedding periods that the databases cover in time. 
The  $R^2$ value of the linear regression exceeds 0.999 in all cases except for the positive vortices in Bu25, which has $R^2>0.99$, confirming the excellent constancy of the advection velocity throughout the investigated domain, $x/D=3$ to $x/D=13$. Table \ref{table2} lists the advection velocity for all three Burger numbers, for vertical planes $z/h=0.12, 0.25, 0.50$. 

\begin{table}
  \caption{Vortex advection velocity ($U_c$) normalised by $U_{\infty}$. Positive and negative vortices are denoted as `$+$' and `$-$', respectively. For all cases, the standard deviation for the advection velocity is less than $5 \times 10^{-3} U_{\infty}$. }
  \begin{tabular}{c  c c c c c c}
    % \hline
    \toprule 
    \multirow{2}{*}{Location} &
      \multicolumn{2}{c}{BuInf} &
      \multicolumn{2}{c}{Bu25} &
      \multicolumn{2}{c}{Bu1} \\
    & $(+)$ & $(-)$ & $(+)$ & $(-)$ & $(+)$ & $(-)$ \\
    % \hline
    \midrule 
    $z/h=0.12$ & 0.877 & 0.880   & 0.894 & 0.888 & 0.902  & 0.910  \\
    % \hline
    \midrule 
    $z/h=0.25$ & 0.889 & 0.890  & 0.877 & 0.885 & 0.901  & 0.914  \\
    % \hline
    \midrule 
    $z/h=0.50$ & 0.868  & 0.872  & 0.868 & 0.889  & 0.891  & 0.916 \\
    % \hline 
    \bottomrule 
    % $z/h=0.75$ & 0.921 & 0.926  & 0.938 & 0.944  & 0.928  & 0.927 \\
    % \hline
  \end{tabular}
  \label{table2}
\end{table}

For BuInf, $U_c$ of positive and negative vortices are practically the same, as expected. Furthermore, $U_c$ exhibits no significant variation from $z/h=0.12$ to $z/h=0.50$. In cases Bu25 and Bu1, anticyclones (positive vorticity) tend to move slower than cyclones (negative vortices), presumably due to their larger size (see figure \ref{fig:omgz_mean}), and this discrepancy is more pronounced at higher planes where vorticity is weaker. Among all the cases, Bu1 has the highest advection velocity as well as the smallest vortex sizes. Nevertheless, in all cases studied here,  regardless of the sign of vorticity, location, and rotation Rossby number, the vortex advection velocities are very close, and can be well approximated by a single value,  $U_c = 0.9 U_{\infty}$. The near constancy of $U_c$ is consistent with the observation of the global modes in section \ref{large-scales} that the tilting angles of the structures do not change as they evolve downstream. It is worth noting that the present results agree well with the measurement by \cite{zhou1992convection} in laminar and turbulent cylinder wakes that the advection velocity is about $0.9 U_{\infty}$. 

% theirs that the advection velocity in the turbulent wakes is found to be around $0.9U_{\infty}$. (Larger than local mean???)

% \jy{cite zhou+antonio and others for advection velocity and Taylors. cite pressure minima + difficulties}. for comparison.  

\begin{figure}[tp]
  \subfloat[]{{\includegraphics[width=.32\linewidth]{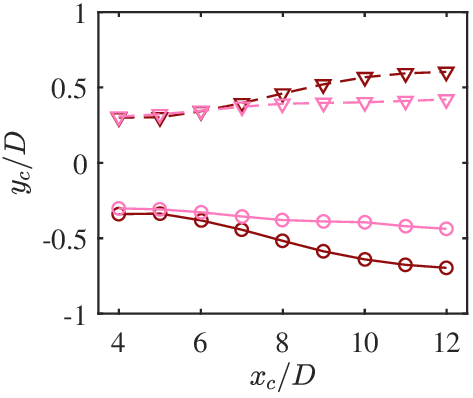}}}% 
  \subfloat[]{{\includegraphics[width=.32\linewidth]{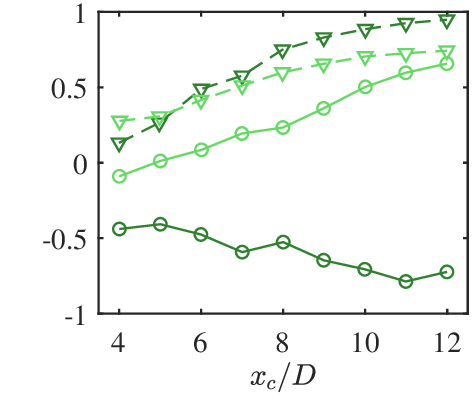}}}% 
  \subfloat[]{{\includegraphics[width=.32\linewidth]{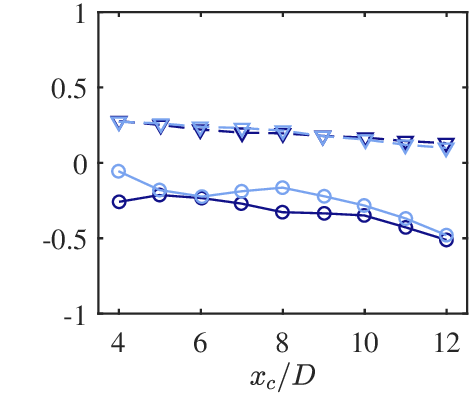}}}% 
  \caption{{Averaged trajectories of identified vortex centres: (a)  BuInf,  (b) Bu25, and (c)  Bu1. Solid lines with circles and dashed lines with triangles represent positive and negative vortices, respectively.  Dark colors are for a lower plane at $z/h=0.12$ and light colors for a higher plane at $z/h=0.50$. }
  % \Sutanu{Break into three subfigs: Buinf, Bu25 and Bu1. It's too busy currently and the information content is difficult to parse! } 
   }
  \label{fig:mean_vor_ctr} 
\end{figure}

The lateral motion of the vortices is also of physical and practical importance. Physically, the lateral movement of the vortices away from the centreline indicates the expansion of the wake and widening of the associated transport of { mass, momentum, and any scalar fields in the flow}. Practically, the mean locations of vortex centres and their variability are instructional to field observations as to where to place measurement stations and to experiments as to where to probe the flowfield. The simple choice of   data sampling on a line at constant $y \neq 0$ is not ideal, as is shown by the curvature of the average path of vortex centres in 
 figure \ref{fig:mean_vor_ctr}. The average is performed locally  {in circles of radius $D/2$ whose successive centers ($x/D=4,5,...,12$) have an  increment of $D$.} Each symbol represents data from all vortex centres that fall in the range of $\pm D/2$ of the $x$-coordinate of the symbol. Solid and dashed lines are the averaged paths of positive and negative vortices, respectively. Results from two planes at $z/h=0.12, 0.50$ are shown. 

{For case BuInf, the distance between positive and negative vortices slightly increases as they are advected downstream, as a result of diffusion of wake vorticity and the expansion of the wake. Taking that as a baseline,  increasing rotation can either widen (Bu25, excluding positive vortices at $z/h=0.50$ where dipoles are formed) or narrow (Bu1) the wake, indicating the nonlinear effect of rotation on wake width growth. } 

% In terms of the distance between cyclones and anticyclones, Bu1 has the narrowest wake, while Bu25 has the widest wake (). Taking the non-rotating case BuInf as the baseline, 

For case Bu1 (figure \ref{fig:mean_vor_ctr} c), the lateral locations of vortex centres are closest to the centreline, compared to the other two cases, and is consistent with the fact that both positive and negative vortices are most compact and smallest at the same $z/h$ location in this case. Moreover, the entire wake characterised by vortex centres is slightly titled to the right ($-y$ direction), in agreement with the direction of the Ekman veering due to an unbalanced pressure gradient at the bottom boundary of rotating flows. {In terms of the vertical alignment, the negative vortices are almost aligned perfectly in vertical throughout the downstream evolution, while positive vortices are not, indicating asymmetry between their properties which will be studied in detail in the next section.}

For case Bu25, the wake is the widest on the left side since the paths of negative vortices have the {largest} deviation from the centreline. 
% \Sutanu{Nice!} 
It is worth mentioning that the light green solid line (at $z/h=0.50$ in Bu25) is special. It represents the path of anticyclonic positive vortices that, statistically speaking,  {do not} reside on the right side ($y<0$) of the hill as {in a typical vortex street configuration,} but deviate  to the left side ($y>0$) instead. {This is due to the formation of vortex dipoles (with uneven vorticity) that translate leftward as viewed from above, as shown by the visualisation in the middle row of figure \ref{fig:omgz_mean}(b). At $Ro=0.75$ (order unity), the shed anticyclonic (positive) vortices are subject to centrifugal-type instability in the near wake and they are consequently larger and weaker, as will be shown in the next section. The insufficient strength of the anticyclones makes them more susceptible to the influence of adjacent cyclones that entrain them to form dipoles. 
% influence of the stronger cyclonic negative vortices on the left side and the resultant formation of vortex dipoles with uneven vorticity that translate leftward, . 
%, and the little distance between positive and negative vortices in figure \ref{fig:mean_vor_ctr}. 
It is worth emphasizing that dipole formation and the resulting leftward deviation of the anticyclones is statistically significant. Generally, anticyclones are expected on the right side of the hill, but this is clearly not true  at  $z/h=0.50$ for the submesoscale Bu25 case with  $Ro = 0.75$.} 

% Each number comes from a least square fit to the extracted trajectories, and the final advection velocity is an average of all trajectories, weighted by their lifetime. 

% $z/h=0.12$ & 0.8906  & 0.8930   & 2.1  & 2.1  & 0.9057  & 0.9185  \\
% \hline
% $z/h=0.25$ & 0.9019 & 0.9045  & 11.6  & 11.6  & 0.9047  & 0.9158  \\
% \hline
% $z/h=0.50$ & 0.9066  & 0.9069  & 5.5  & 5.5  & 0.8989  & 0.9237 \\
% \hline 
% $z/h=0.75$ & 0.9211 & 0.9264  & 5.5  & 5.5  & 0.9280  & 0.9274 \\

% Bu1_vor_stat_i0020

% For upper planes, rotation is less felt and adv vel is similar for bu1 and Buinf. For bu 1, lower 3 planes, cyclones are faster and smaller compared to anticyclones, statistically. 

% Turbulence level is low. Structures organise into coherent structures quickly. Mean deficit decay is slowed possibly as a result.

% it might also be of interest to the investigation of other flows where the tracking of individual vortex centres is useful. 

% \jy{Mittal has a paper (2nd street?) saying the adv. vel. is a constant. Compare to them.} 

% Antonia, Zhou. 

% Surprisingly, total number of graphs is quite uniform, being between 80-85 for all cases and locations. 

% \cite{chopra1965mesoscale} found the advection speed for the vortices in an island wake is about $85\%$ of the freestream velocity. 
% Since we knew that the adv vel is roughly a constant during the evolution, we don't distinguish different streamwise locations when doing the conditional average. 

  %%%%%%%%%%%%%%%%%%%%%%%%%%%%%%%%%%%%%%%%%%%%%%%%%%%%%%%%% 
\section{Cyclones and anti-cyclones; marginal instability} 
\label{cyclones}

In non-rotating unsteady wakes of bluff bodies with a symmetrical cross-section, there exists no statistical asymmetry {in the mean} between positive or negative vorticity as the reflectional symmetry is respected. As a result, in a classic cylinder wake or the present non-rotating case (case BuInf), the mean vertical vorticity is antisymmetric with respect to the centreline $y=0$ (see figure \ref{fig:omgz_mean}d). However, the  Coriolis force that accompanies system rotation breaks this reflectional symmetry. 
%\st{Moreover, vortices that rotate in the same direction as the reference frame rotation (referred as cyclonic vortices, or CVs) are preferred over those rotate in the opposite direction (anticyclonic vortices, AVs).}  \Sutanu {Preceding crossed out statement is not generally true!} \jy{jy: By `preferred' I meant stability in this particular obstacle-generated vortices case. Thank you for being confused and for the papers.} 
As the Coriolis frequency ($f_{\rm c}$) is negative in this study, positive vortices ($\omega_z>0$) are AVs and vice versa. The CVs and AVs present considerable differences in the rotating cases as illustrated by the visualisations of figure \ref{fig:omgz_mean}. 

This section is arranged as follows: section \ref{prob} compares the probability distribution function (p.d.f.) of positive and negative $\omega_z$ in different cases and examines the systematic asymmetries and biases that rotation introduces to vorticity; section \ref{cond} utilises the vortex centres extracted in the previous section to obtain ensemble-averaged conditional statistics to characterize how vortex structure  depends on rotation; \ref{stab} elaborates on the stability properties of AVs and their implication. 

\subsection{Probability distribution of vorticity} \label{prob}

\begin{figure}[tp]
  \subfloat[]{{\includegraphics[width=.5\linewidth]{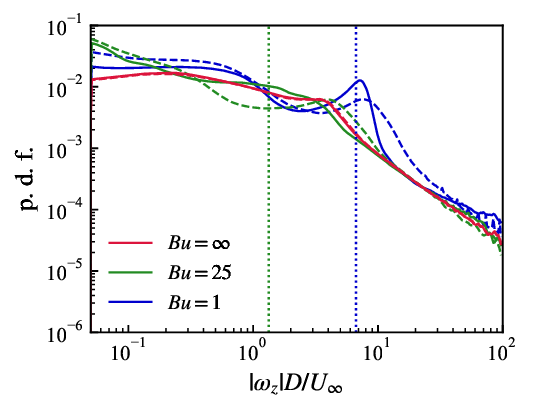}}}% 
  \subfloat[]{{\includegraphics[width=.5\linewidth]{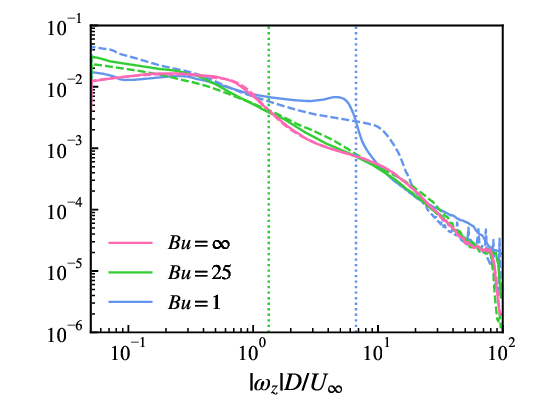}}}% 
  \caption{P.d.f. of $|\omega_z|$ at (a) $z/h=0.12$, and (b) $z/h=0.50$. Solid lines indicate positive vorticity (AVs), and dashed lines indicate negative vorticity (CVs). {Vorticity is presented in convective units and the dotted vertical reference lines represent the magnitudes of the non-dimensionalised Coriolis frequency $|f_{\rm c}| D/U_{\infty}$, which equals $1/Ro$ and takes the value of $4/3$ for case Bu25 and $20/3$ for case Bu1, respectively.} } 
  \label{fig:pdf} 
\end{figure}

The p.d.f. of  $|\omega_z|$ for positive vorticity (solid line) and negative vorticity (dashed line) are shown separately and on the planes $z/h=0.12$ (figure \ref{fig:pdf}a)  and  $0.50$ (figure \ref{fig:pdf}b). Each p.d.f. is  normalised such that the area under each line is equal. {In the p.d.f., $|\omega_z|$ is normalised with convective units ($D$ and $U_{\infty}$) for consistency between rotating and non-rotating cases (where $f_{\rm c}$ is zero in BuInf), but the Coriolis frequency  ($|f_{\rm c}|$)  will also be used for normalization in cases Bu25 and Bu1.}
% to represent the relative magnitude of $\omega_z$.
  For case BuInf, symmetry is achieved for vorticity of all magnitudes, as expected. Comparing cases Bu25 and Bu1, there is an increasing asymmetry between the p.d.f. of positive and negative vorticity, as the rotation strength is increased. 

Consider the Bu1 case (blue lines). In plane $z/h=0.12$ shown in figure \ref{fig:pdf}(a), {there is a  local peak for cyclonic vorticity (negative $\omega_z$) and one for anticyclonic vorticity (positive $\omega_z$),  both being slightly larger than the system rotation rate and close to $1.1|f_{\rm c}|$. In plane $z/h=0.50$ shown in figure \ref{fig:pdf}(b), there is a local peak only for anticyclonic (positive) vorticity at $0.7 |f_{\rm c}|$.}
%\st{The  former is an evidence of the enhancement of cyclonic vorticity by rotation. To show preceding don't we have to compare the un-normalized pdf values?} \jy{jy: revised. actually $\omega_z/f_c$ is higher in Bu25 than Bu1 shown in latter section or OSM abstract}. 

For case Bu25, the p.d.f. of cyclonic vorticity {at $z/h=0.12$ shown in figure \ref{fig:pdf}(a) also has a peak above $|f_{\rm c}|$ (with peak relative vorticity $ 3.1 |f_{\rm c}|$, which is significantly greater that in case Bu1).} On the other hand,  anticyclonic vorticity does not show any observable peak near $|f_{\rm c}|$.
% \st{ suggesting the breaking of those vortices due to instability and hence a more widespread distribution of $\omega_z$.} \Sutanu{ We can't infer preceding statement.}

% {In plane $z/h=0.50$  shown in figure \ref{fig:pdf}(b), the general shift of the p.d.f. to the left indicates that vorticity is weaker on average as the local hill diameter is smaller. For case Bu1, the local peak is only observed for anticyclonic vorticity at a value that is weaker than $|f_{\rm c}|$, suggesting the effect of centrifugal adjustment. However, in case Bu25, there is not much asymmetry between CVs and AVs in the plane $z/h=0.50$ compared to $z/h=0.12$, although,  the vorticity distributions inside CVs and AVs are very different, which will be a focus of the next section. }

The local peaks in p.d.f.s are interpreted as the values that are more commonly found in the flow compared to their neighbours, instead of the most intense ones. In the next section, we will show the existence and persistence of intense anticyclones ($\omega_z > |f_{\rm c}|$) in both Bu1 and Bu25 and discuss their stability.

%\jy{The near- and sub-inertial peaks of anticyclonic vorticity in case Bu25 and Bu1 correspond to a state of stable absolute vorticity ($(\omega_z + f_{\rm c})f_{\rm c} \gtrapprox 0$) that is often found in the anticyclones owing to the cyclo-geostrophic instability. However, the local peaks in p.d.f.s are interpreted as the values that are more commonly found in the flow compared to their neighbours, instead of the most intense ones. In next section, we will show the existence and persistence of intense anticyclones ($\omega_z > |f_{\rm c}|$) in these two rotating cases and discuss their stability. }  

% At the two heights, both cyclonic and anticyclonic vorticity are higher in terms of convective unit ($U_{\infty}/D$) in case Bu1 showing the strong influence of rotation in intensifying vorticity.

\subsection{{Vorticity conditioned to} individually tracked vortices} \label{cond}

The vorticity distribution inside wake vortices is essential to the understanding of their kinematics, the idealisation and modelling of such vortical wakes, and the role of vortex stability. {The previous section on vorticity p.d.f., while  giving an overall statistical view of  cyclonic/anticyclonic vorticity, does not reveal the properties of individual coherent wake vortices -- the focus of this section.} The identification of vortex centers in  section \ref{tracking}  is leveraged to quantify the  vorticity conditioned to the identified vortex centres and thus  reveal the {flow inside and around} the vortices.  {As elaborated below, rotation significantly impacts the downstream evolution of the vortex-conditioned distribution of $\omega_z$ and, notably, the difference between AV and CV is larger for $Bu = 25$ than for $Bu =1$.}

\begin{figure}[pt]\captionsetup[subfloat]{farskip=-12pt,captionskip=1pt}
  \subfloat[]{{\includegraphics[width=.45\linewidth]{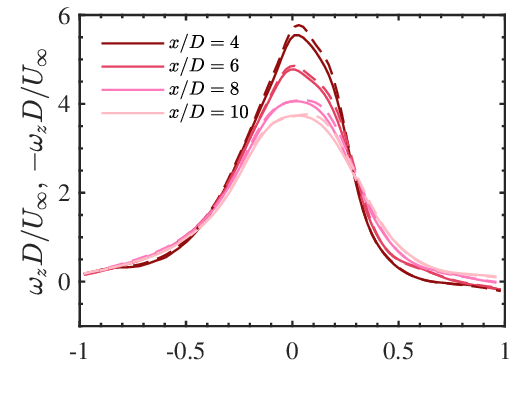}}}% 
  \subfloat[]{{\includegraphics[width=.45\linewidth]{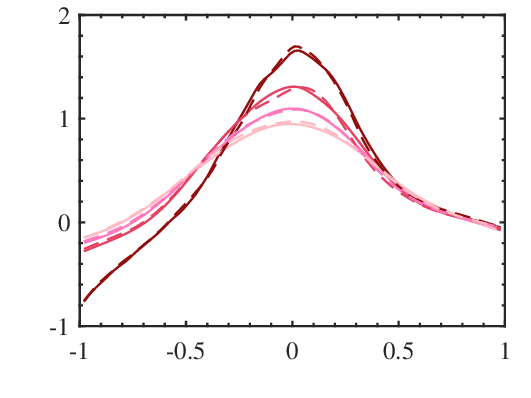}}}% 
  \\ 
  \subfloat[]{{\includegraphics[width=.45\linewidth]{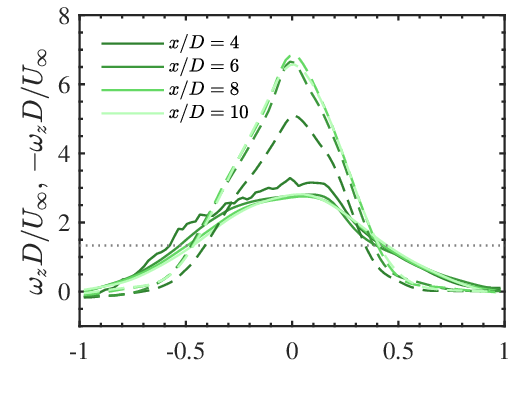}}}% 
  \subfloat[]{{\includegraphics[width=.45\linewidth]{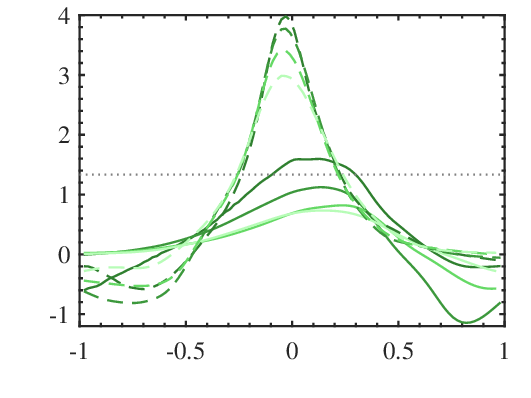}}}% 
  \\
  \subfloat[]{{\includegraphics[width=.45\linewidth]{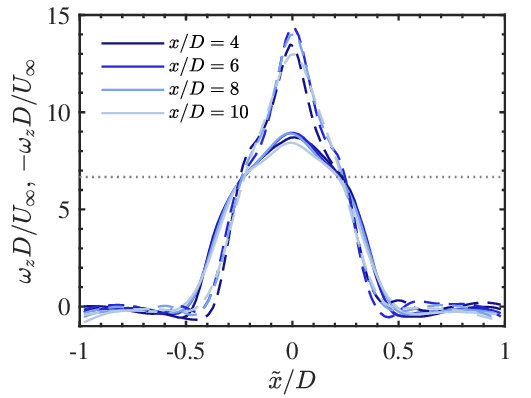}}}%  
  \subfloat[]{{\includegraphics[width=.45\linewidth]{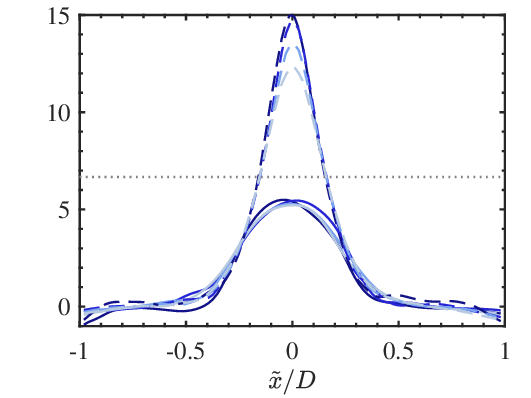}}}% 
  \\ 
  \caption{Conditionally averaged vorticity  {distribution around vortex centres, $\omega_zD/U_{\infty}$ for} positive vortices (AVs, solid) and  {$-\omega_zD/U_{\infty}$} for negative vortices (CVs, dashed). (a,b) BuInf, (c,d) Bu25, and (e,f) Bu1. Horizontal planes shown are at $z/h=0.12$ (a,c,e) in the left column and at $z/h=0.50$ (b,d,f) in the right column. The horizontal dotted lines in (c,d,e,f) indicates the  {case-dependent} absolute magnitude of the Coriolis frequency ($|f_{\rm c}|$). In plots (c-f), the strongest CVs have peak magnitudes of {$\omega_z/|f_{\rm c}|=5.1, 3.0, 2.2, 2.2$}, and the strongest AVs have peaks {$\omega_z/|f_{\rm c}|=-2.4, -1.2, -1.3, -0.8$}, respectively.}
  \label{fig:vor_ctr_stat} 
\end{figure} 

Figure \ref{fig:vor_ctr_stat} shows  profiles of the average of $\omega_z(\tilde{x})$, conditioned to instantaneous individual vortex centres. {Here, $(\tilde{x},\tilde{y})=(x-x_c,y-y_c)$ is a new set of horizontal coordinates fixed to individual vortex centres.} 
% \st{Here, $\tilde{x}= x -x_c$ with $x_c$ denoting the streamwise location of the vortex center.} 
% \st{Better to introduce a new coordinate, $x^*= x -x_c$ so that  $\omega_z(x^*)$ is shown for various $x_c$.} 
  Since wakes are spatially developing, the results are presented at various values of $x_c$  to diagnose the  downstream evolution of vortex-conditioned properties. Vortices with centres apart less than $2D$ are assumed to possess similar properties and grouped for a regional average. For example, the group located at {$x_c/D=4$} represents vortex centres that fall in the section of $3<x/D<5$ and so forth. Each vortex centre in the same group is shifted to {$\tilde{x} = \tilde{y}=0$} before the statistics are gathered. For each group, more than 2000 vortices are available for the ensemble average.  

%\Sutanu{A question that reviewers might have is why not look at $\omega_z(y)$, too? } 
%\jy{jy: the answer is the result will be (quantitatively) different but qualitative the same since everywhere is stable. and we are not supposed to do azimuthal average since vortices are not circular at least in Buinf.}

For case BuInf shown in figure \ref{fig:vor_ctr_stat}(a,b), vorticity profiles are Gaussian-like except close to the edges.  The  peak magnitude decays and the width grows  as a result of diffusion. Otherwise, no significant change  is present when $x$ increases. In figure \ref{fig:vor_ctr_stat}(a) at $z/h=0.12$, the magnitude of $\omega_z$ near $x/D=\pm 1$ is close to zero -- the asymptotic far-field condition for isolated vortices. However, in figure \ref{fig:vor_ctr_stat}(b) at $z/h=0.50$, the vorticity can change sign when $\tilde{x}/D$ varies between $0$  and $-1$, becoming substantially negative for the group $x_c/D=4$ but less so further downstream. A similar sign change is not observed when $\tilde{x}$ varies between $0$ and $1$. As seen in the middle and bottom rows of figure \ref{fig:omgz_mean}(a),  wake vortices of both signs at $z/h=0.50$ are more spatially diffuse than at $z/h=0.12$ and are almost side-by-side in the region around $x/D=4$, making possible  the sign change of vorticity.
%it is likely when $x/D$ is small and there is not enough lateral separation between opposite-sign vortices. 

For case Bu25 shown in figure \ref{fig:vor_ctr_stat}(c,d), CVs and AVs are substantially different. CVs are stronger and more compact, while AVs are weaker and wider. The latter is likely because of the cyclo-geostrophic instability associated with the AVs upstream before the conditional statistics are gathered. In figure \ref{fig:vor_ctr_stat}(c), the vorticity distribution for $x_c/D=4$ displays short-wavelength wiggles and represents active instability of the AVs as can also be seen in the bottom row of figure \ref{fig:omgz_mean}(b). { We emphasize that the short-wavelength wiggles are in a profile that is obtained by averaging over an ensemble of approximately 2000 members and are, thus,  statistically significant.}  This is later confirmed in section \ref{stab} where the generalised Rayleigh discriminant on the left side of the aforementioned AVs is shown to lie beyond the stability limit.

Figure \ref{fig:vor_ctr_stat}(d) shows differentiated behavior between CVs and AVs when the vortex edge is approached. The conditionally averaged vorticity in  AVs (positive $\omega_z$) tends to change sign toward the right end, while that around CVs (negative $\omega_z$) changes sign toward the left end. This indicates a preferred configuration of vortex dipoles with a positive vortex on the left and a negative vortex on the right, as can {also} be seen in the middle row of figure \ref{fig:omgz_mean}(b). At the same time, the dipoles are quite asymmetric; the positive $\omega_z$ AV is weaker and more spatially diffuse than its partner, suggesting that the weaker AVs are more susceptible to the induced motion of the CVs.  {It is also in case Bu25 that the asymmetry between CVs and AVs is most pronounced. As shown in} \ref{fig:vor_ctr_stat}(d), {the ratio of peak magnitudes between cyclones and anticyclones is approximately four on average, and increases as downstream distance is increased. In combination with the fact that the absolute magnitude of the anticyclones is weak (order $D/U_{\infty}$), they are more easily influenced by adjacent cyclones during mutual interaction.}

Strong rotation favors the formation of coherent vortices. For case Bu1 shown in figure \ref{fig:vor_ctr_stat}(e,f), both CVs and AVs are strongest in units of $U_{\infty}/D$ in all three cases. In the plane at $z/h=0.12$, both CVs and AVs have magnitude greater than $|f_{\rm c}|$ while only CVs exceed $|f_{\rm c}|$ in the plane at $z/h=0.50$. Moreover, both CVs and AVs undergo little change in terms of vorticity magnitude and distribution during their advection --  a key difference from the other two Burger numbers.
This is in agreement with the mean flow characteristics to be discussed in section \ref{statistics}  that the  streamwise change in the momentum wake of Bu1 is slower than in the other two cases. 

In terms of the strength of the AVs, the vorticity magnitude in the vortex core exceeds $|f_{\rm c}|$ at all streamwise locations in the plane at $z/h=0.12$, {in both cases Bu25 and Bu1}.  Thus, the absolute vorticity stability criterion is not conclusive to the behavior of AVs  since, contrary to that stability  criterion, AVs with $|\omega_z| > f$ are found to advect in a stable manner.

% \Sutanu{I think a table with subtables would be useful as a quantitative birds-eye view (a potential format for subtable follows) 
% \begin{equation}
% {\rm Case}  \quad   \omega_{p,CV} \quad \omega_{p,AV}    \quad l_{0.1,CV}  \quad l_{0.1,AV}
%  \end{equation} 
%  Each vorticity column with 2 subcolumns, one in convective and the other in $f$ units. Three cases: Buinf, Bu25 and Bu1. $l_{0.1}$ is with based on magnitude of 0.1 times peak value, small enough to capture entire vorticity distribution (excluding sign change for Bu = 25, $z/h = 0.5$)
 
%  Say, a 2 x 2 matrix of subtables at the two planes and at $x_c/D = 4$ and $x_c/D = 10$.\\
%  }

\begin{table}[tbh]
    \centering
    % \aboverulesep=0ex % Solution part 1 of 3
    % \belowrulesep=0ex % Solution part 1 of 3
    \caption{{Summary of the properties of cyclonic and anticyclonic vortices, at two horizontal planes $z/h=0.12, 0.50$ and two streamwise locations of vortex centres $x_c/D=4, 10$. The peak vertical vorticity $\omega_{z,p}$ is given in convective ($U_{\infty}/D$) as well as rotation ($f_{\rm c}$) units. Vortex sizes are characterised by $l_{0.05}$, which is the diameter at which the intensity of $\omega_z$ decays to $0.05\omega_{z,p}$.}}
    \renewcommand\multirowsetup{\centering}
    \begin{tabular}{ c c c c c c c c c }
    \toprule 
                          &                     &                          & \multicolumn{3}{c }{CV}  & \multicolumn{3}{c}{AV}   \\
    % \cline{2-3}
    \multirow{-2}{*}{case}  &  \multirow{-2}{*}{$z/h$} &  \multirow{-2}{*}{$x_c/D$}     &     $-\omega_{z,p}D/U_{\infty}$  & $-\omega_{z,p}/|f_{\rm c}|$ & $l_{0.05}/D$         & $\omega_{z,p}D/U_{\infty}$  & $\omega_{z,p}/|f_{\rm c}|$ & $l_{0.05}/D$ \\
    % \hline
      \midrule 
    \multirow{4}{*}{Bu25} & \multirow{2}{*}{0.12} &  4  & 5.1  & 3.8 & 1.09 & 3.2 & 2.4 & 1.64 \\ 
                          &                       &  10 & 6.6  & 4.9 & 1.15 & 2.8 & 2.1 & 1.73 \\ \cmidrule[0.5pt]{2-9} % 
                          & \multirow{2}{*}{0.50} &  4  & 4.0  & 3.0 & 1.15 & 1.6 & 1.2 & 1.23 \\
                          &                       &  10 & 3.0  & 2.2 & 1.02 & 0.7 & 0.6 & 1.45 \\ \midrule % 
    \multirow{4}{*}{Bu1}  & \multirow{2}{*}{0.12} &  4  & 13.5 & 2.0 & 0.74 & 8.7 & 1.3 & 0.88 \\ 
                          &                       &  10 & 13.0 & 2.0 & 0.80 & 8.4 & 1.3 & 0.93 \\ \cmidrule[0.5pt]{2-9}
                          & \multirow{2}{*}{0.50} &  4  & 15.0 & 2.3 & 0.75 & 5.5 & 0.8 & 0.80 \\
                          &                       &  10 & 12.3 & 1.9 & 0.82 & 5.2 & 0.8 & 1.04 \\
    % r1 & a & r1 & a & b & r1 & a & b & a  \\
    \bottomrule 
    % r2 & c & d   \\
    % \hline
    \end{tabular}
    \label{table3}
\end{table}

{Table \ref{table3} summarises  the properties of CVs and AVs, in terms of peak intensity and vortex size. The peak intensity $\omega_{z,p}$ is the maximum of $\omega_z$ of each curve in figure \ref{fig:vor_ctr_stat}, and the vortex size $l_{0.05}$ is the horizontal distance within which the magnitude of $\omega_z$ is  greater than $0.05 \omega_{z,p}$. It can be seen that at the same spatial location, vortex intensity is generally higher in Bu1 in convective units ($U_{\infty}/D$).
%but is higher in Bu25 in rotation unit ($f_{\rm c}$). The only exemption is AVs at $z/h=0.50$ and $x/D=10$. This is again showing the non-proportional effect of rotation in intensifying vorticity in the same direction. 
Moreover, the sizes of both CVs and AVs are consistently smaller in case Bu1. In combination with greater peak vorticity, velocity gradients are much larger in vortices in case Bu1 than in Bu25.
In terms of vorticity in units of $f_{\rm c}$, Bu25 stands out with the largest magnitude of $\omega_z/|f_{\rm c}|$.}

% \begin{table}[t!]
%   \caption{ }
%   \begin{tabular}{c c c c c c c c c}
%     \hline 
%     \multirow{2}{*}{case}   & \multirow{2}{*}{plane($z/h$)}   & \multirow{2}{*}{streamwise location ($x/D$)}  & 
%     \multicolumn{3}{*}{CV}  & \multicolumn{3}{*}{AV} \\
%                             &                                 &                                               & 
%          wp & wp & l        &     wp & wp & l         
%     % \multicolumn{3}{*}{CV} & \multicolumn{3}{*}{AV} \\ 
%     % \hline 
%     % Case &$Bu$&$Ro$ & $Fr$                  & $(N_x,N_y,N_z)$ & $N_t$ & $T U_{\infty}/D$ & color code \\ 
%     % \hline 
%     % BuInf  &$\infty$   &$\infty$& \multirow{3}{*}{0.15} & \multirow{3}{*}{(1536,1280,320)} &  \multirow{3}{*}{4000} & 295  & red\\ 
%     % Bu25 & 25       & -0.75      &          &  & & 345 & green\\
%     % Bu1& 1 & -0.15 &          &  & & 322 & blue \\
%     % \hline 
%     \hline 
%   \end{tabular} 
%   \label{table3}
% \end{table} 

% Even the rotation in Bu25 is able to change the vertical structures from slanted `sheets' to `columns', it makes a difference in the instability of anticyclones, 

% We have already shown that the turbulence level is low in the wake and the transition to three-dimensional turbulence is discouraged. Hence, it is suitable to consider stability of anticyclones in our cases. 

% \begin{figure}
%   \missingfigure{velocity distribution inside vortices.} 
% \end{figure} 

\subsection{Stability of anticyclones} \label{stab}
% \st{Oceanic and atmospheric flows involve multiple physical mechanisms that could potentially influence  the stability of the eddies}. \Sutanu{Let's avoid preceding since most of these mechanisms are associated to topographic interaction,  upper-ocean processes, internal waves, etc. which we do not address. }

{Vortex stability is an important question since it is  related to the generation of turbulence and small-scale motions. Here, we assess the ability of various criteria in the literature to identify the stability characteristics of the advecting wake vortices of the present simulations. It will be shown that more recent criteria that account for stratification and viscous dissipation constitute a significant improvement over earlier attempts. }

%However, with the interplay among rich physics, it becomes challenging to establish a sufficient and necessary condition for instability. 
The study of the centrifugal (inertial) instability of swirling flows can be traced back to \cite{rayleigh1917dynamics} who showed that the flow could become unstable if the squared angular momentum decreases radially. The Rayleigh criterion is a necessary and sufficient condition for the instability of inviscid columnar vortices subject to three-dimensional axisymmetric perturbations \citep{drazin2002introduction}. In a rotating frame, the Rayleigh criterion for inertial instability is equivalent to the existence of a region with a negative product of Coriolis frequency and absolute vorticity \citep{holton1972introduction}, 

\begin{equation}
  {f_{\rm c}}({\omega_z (r) + f_{\rm c} }) < 0, \label{abs_vor}
\end{equation}
or in terms of the local Rossby number  {(note $f_{\rm c}$ can take either sign)},
 \begin{equation}
  {Ro_{\rm L} (r)}  = \frac{\omega_z (r) }{f_{\rm c}} < -1 . \label{RoRayleigh}
\end{equation} 
The  {criterion for inertial instability},  (\ref{abs_vor}),  is  widely used and implies that  anticyclones with vorticity magnitude exceeding  {$|f_c|$} are unlikely.
% However, the criterion \eqref{abs_vor} was established by energy arguments  \hl{similar to} \cite{rayleigh1917dynamics} \hl{Why the preceding phrase?}, with a unidirectional geostrophic wind assumed as the base state \citep{holton1973introduction}. 
{However, it assumes a sheared parallel flow as the base state \citep{holton1972introduction}.}
\cite{kloosterziel1991experimental} found in their experiment that CVs could be unstable too, {contrary to  (\ref{RoRayleigh}) }.
With the inclusion of the centrifugal term, a generalised Rayleigh discriminant \citep{kloosterziel1991experimental,mutabazi1992gap} for centrifugal instability in rotating vortical flows was established as 
\begin{equation} 
  \chi(r) = [\omega_z(r)+f_{\rm c}][2\frac{u_{\theta}(r)}{r}+f_{\rm c}] < 0,  \label{gen_ray}
\end{equation} where $\omega_z+f_{\rm c}$ is interpreted as the absolute vorticity and $u_{\theta} + f_{\rm c} r/2$ as the absolute velocity. It implies instability when the absolute velocity and absolute vorticity are of opposite signs. Equation \eqref{gen_ray} assumes a specific form (axisymmetric)  of perturbations.  Nevertheless,  \eqref{gen_ray}  is good enough in many circumstances since axisymmetric perturbations are, in general, more unstable than non-axisymmetric ones \citep{billant2005generalized}.

%Apart from the inviscid cyclo-geostrophic instability mentioned above 
{While the former two criteria \eqref{abs_vor} and \eqref{gen_ray} concern the stability to 2D perturbations, the inclusion of effects in the third dimension, such as stable stratification, finite dissipation, and the vertical variation of the  vortex base flow, is necessary for actual geophysical flows and will complicate the determination of stability}. With stratification that suppresses small wavenumbers and finite vertical dissipation that suppresses large wavenumbers, the range of unstable vertical wavenumbers is reduced and
 %he sufficient condition 
 \eqref{gen_ray}
  %becomes a necessary condition as it
   overestimates the unstable region \citep{lazar2013inertiala}. To reduce the predicted unstable region, \cite{lazar2013inertiala} proposed a new criterion for the marginally stable Burger number: 
\begin{equation} 
  \sqrt{Bu_{\rm v}} = \left(\frac{3}{8|a_0|}\right)^{3/2} \frac{1}{\sqrt{Ek}} \frac{(|2Ro_{\rm v}+1|)^{7/4}}{|Ro_{\rm v}|}, \label{lazar}
\end{equation} where \begin{equation}
  Ro_{\rm v}= \frac{V_{\rm max}}{f_{\rm c} r_{\rm max}} ; \, Bu_{\rm v} = \left(\frac{Nh}{f_{\rm c} r_{\rm max}}\right)^2 ; \,  Ek = \frac{\nu}{{|f_{\rm c}|} h^2}
  \label{vor_bur}
\end{equation}
are the vortex Rossby number, the vortex Burger number, and the vertical Ekman number. Here, $V_{\rm max}$ and $r_{\rm max}$ refer to the peak  {magnitude} of the azimuthal velocity and its location, respectively.  {Positive and negative $Ro_{\rm v}$ represent cyclones and anticyclones, respectively. The constant $a_0=-2.338$ is the first zero of the Airy function.} In the parameter space of $Bu_{\rm v}$-$Ro_{\rm v}$, Eq. \eqref{lazar} at each constant $Ek$ corresponds to a stability curve that separates regimes that are  stable and unstable to axisymmetric perturbations. 

{More recently}, \cite{yim2019stability} pointed out that the most unstable azimuthal wavenumber of the centrifugal mode is not necessarily $m=0$ (axisymmetric), but depends on $Bu_{\rm v}$ {and will have an impact on the determination of stability}. Accordingly, they suggested  the use of the stability curve {given as}
 \begin{equation}
  \sqrt{Bu_{\rm v}} = \frac{0.23}{\sqrt{Ek}} \frac{(Ro_{\rm v}+0.3)^2}{\sqrt{|Ro_{\rm v}|}},  \label{yim} 
\end{equation}
{which considered the dependence of the most unstable azimuthal wavenumbers on ${Bu_{\rm v}}$.} {It was also found that the stability of AVs is insensitive to the vertical variation of the initially axisymmetric vortex base flow.}  
%  However,  \cite{lazar2013inertiala,yim2019stability} suggested new stability criterions of AVs in rotating-stratified moderate Reynolds number flows, both involving the vertical Ekman number, as well as vortex Rossby and Burger numbers. 
In this section, we will compare the absolute vorticity criterion \eqref{abs_vor}, the generalised Rayleigh discriminant \eqref{gen_ray}, as well as the new criteria \eqref{lazar} and \eqref{yim}, 
% proposed by \cite{lazar2013inertiala} and \cite{yim2019stability}, respectively, 
% in moderate-to-high Reynolds number stratified flows, 
 {and assess} their ability to predict the stability of the wake vortices in the simulations.

\begin{figure}[pt]\captionsetup[subfloat]{farskip=-12pt,captionskip=1pt}
  \subfloat[]{{\includegraphics[width=.45\linewidth]{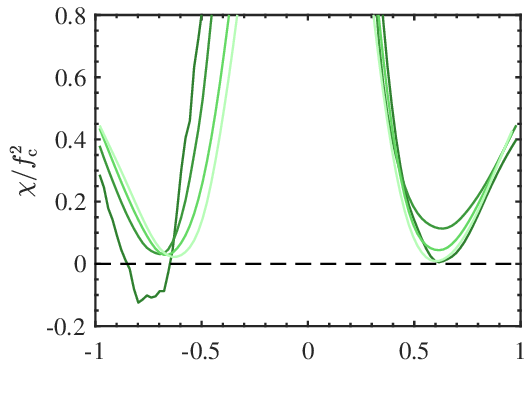}}}% 
  \subfloat[]{{\includegraphics[width=.45\linewidth]{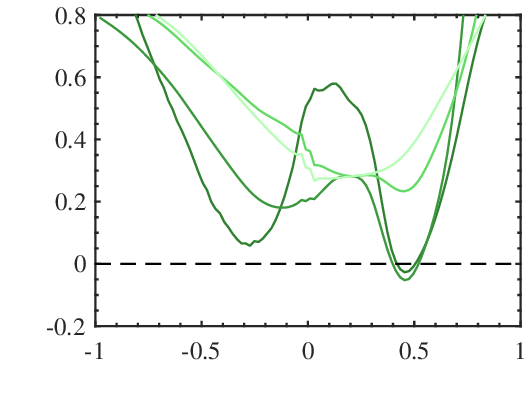}}}% 
  \\ 
  % \subfloat[]{{\includegraphics[width=.45\linewidth]{./figs/Bu25_chi_i0038.eps}}}% 
  % \subfloat[]{{\includegraphics[width=.45\linewidth]{./figs/Bu1_chi_i0038.eps}}}% 
  % \\
  \subfloat[]{{\includegraphics[width=.45\linewidth]{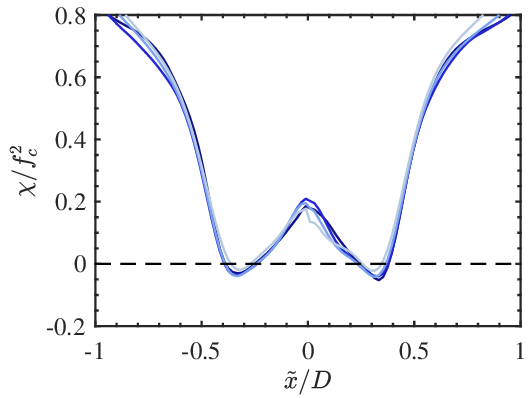}}}%  
  \subfloat[]{{\includegraphics[width=.45\linewidth]{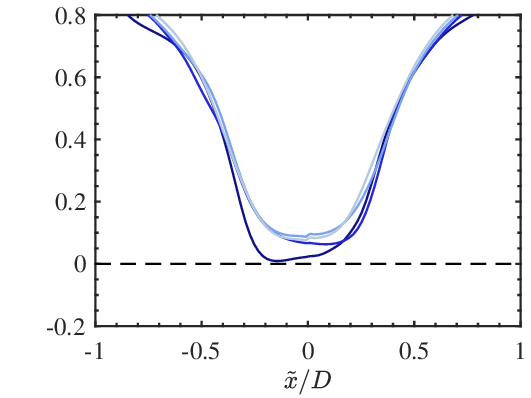}}}% 
  \\
  \caption{Conditionally averaged generalized Rayleigh discriminant $\chi/f_{\rm c}^2$ for AVs in cases Bu25 (a,b) and Bu1 (c,d). Horizontal planes shown are at $z/h=0.12$ (a,c), and at $z/h=0.50$ (b,d). Streamwise locations, $x/D = 4, 6, 8$ and 10 are shown in  dark to light colors. The dashed line in each figure marks the stability criterion of $\chi = 0$.}
  \label{fig:chi_stat}  
\end{figure}

As was shown in figure \ref{fig:vor_ctr_stat}(c,e), anticyclones at $z/h=0.12$ are observed to possess a large region with $|\omega_z|>|f_{\rm c}|$, although they advect as stable vortices for a significant range of downstream evolution distance. Hence, the absolute vorticity condition \eqref{abs_vor} is not sufficient for instability. 

Figure \ref{fig:chi_stat} shows the conditionally averaged generalized Rayleigh discriminant $\chi(\tilde{x})$ in \eqref{gen_ray} as a function of streamwise distance from the vortex centre. It can be seen that the unstable region is reduced significantly compared to \eqref{abs_vor}. In case Bu25 and at $z/h=0.12$ (figure \ref{fig:chi_stat}a), at roughly $x/D=4$ (the darkest green line), there is a small region, located near the left edges of the AVs, that is found to be unstable, but otherwise, all regions have at least marginally stable $\chi$. Furthermore, the vortices tend to evolve to a more stable state. For Bu1, the AVs in the $z/h=0.50$ plane (figure \ref{fig:chi_stat}b) have stable discriminant $\chi$, while the peripheries of the AVs in the plane $z/h=0.12$ have marginally unstable $\chi$ which does not experience noticeable change during the advection. The statistics of $\chi$ of the AVs in the plane at $z/h=0.25$ are similar to those at $z/h=0.12$ and are not shown. Consistent  with the instantaneous snapshots in figure \ref{fig:omgz_mean} and the statistics in figure \ref{fig:vor_ctr_stat},   AV instability is only observed for Bu25 at $z/h=0.12$, and is captured properly by the generalized Rayleigh discriminant \eqref{gen_ray}. On the other hand, in case Bu1 at $z/h=0.12$ (figure \ref{fig:chi_stat}c) where the discriminant is marginally unstable at the periphery of the vortices, no actual change to the vorticity profile of the AVs is observed (see figure \ref{fig:vor_ctr_stat}e). Hence, 
%this marginal instability might still be falsely identified as the generalized Rayleigh discriminant becomes a necessary condition for instability 
a sufficient condition for stability requires other considerations, e.g., stratification and dissipation \citep{lazar2013inertiala,yim2019stability}. 

% $V(x)$ is similar to those in \cite{lazar2013inertialb} partly justifying substituting $r$ with $x$. 

\begin{figure}[pt]\captionsetup[subfloat]{farskip=-12pt,captionskip=1pt}
  \subfloat[]{{\includegraphics[width=.45\linewidth]{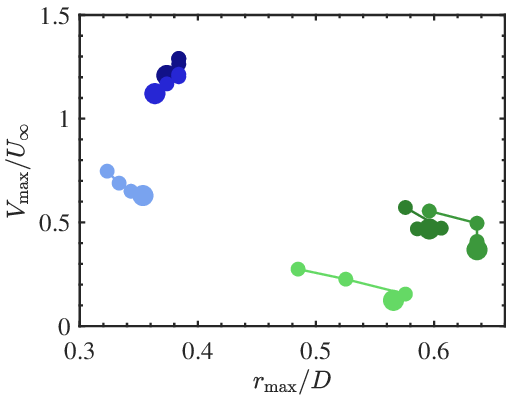}}}% 
  \subfloat[]{{\includegraphics[width=.45\linewidth]{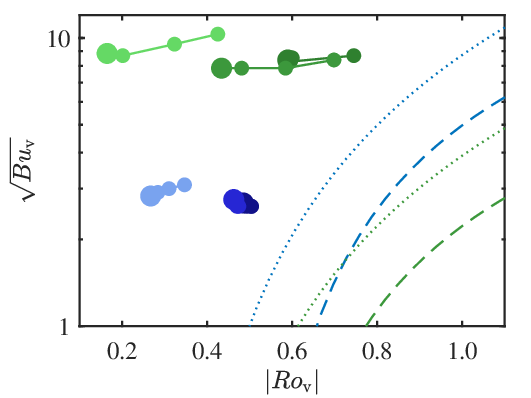}}}% 
  \\ 
  \caption{Vortex properties: (a) conditionally averaged maximum azimuthal velocity $V_{\rm max}$ and  corresponding location $r_{\rm max}$ and (b) square-root of the vortex Burger number $\sqrt{Bu_{\rm v}}$ and  the vortex Rossby number $Ro_{\rm v}$.  Bu25 is shown in green and Bu1 in blue. Colors from dark to light indicate planes at $z/h=0.12, 0.25, 0.50$. Each circle marks one of the locations from $x/D=4,6,8,10$.  {The circles are connected by lines following the order of $x$-locations} and the last location ($x/D=10$) is marked with the largest circle. In (b), dashed lines denote  the stability curves \eqref{lazar} by \cite{lazar2013inertiala} and dotted lines \eqref{yim} by \cite{yim2019stability}  for cases Bu1 (blue) and Bu25 (green). 
  {The Ekman numbers are  $Ek=|Ro|/Re_D (D/h)^2 = $ $ 8.33 \times 10^{-4}$ and $ 1.67 \times 10^{-4}$ for cases Bu25 and Bu1, respectively.} 
  % \Sutanu{The values of $Ek$ are ?? for Bu1 and ?? for Bu25.}  
  The left side of the neutral stability curves is stable. }
  \label{fig:para_space} 
\end{figure} 

%In order to better infer the AV stability, 
% To {obtain $Ro_{\rm v}$ in} \eqref{vor_bur},  
 {Prior to applying \eqref{lazar} and \eqref{yim},} 
the vortex sizes and shapes in terms of $V_{\rm max}$ and $r_{\rm max}$ are required. The radial direction is substituted by the streamwise ($x$) direction and the azimuthal velocity component  by the spanwise velocity ($v$). The peak velocity $V_{\rm max}$ is defined as the maximum azimuthal (herein transverse) velocity and the corresponding peak location $r_{\rm max}$ is interpreted as vortex radius. {It is noted that for both criteria \eqref{lazar} and \eqref{yim} applied here, 2D vortex profiles subject to 3D perturbations are assumed, and the vertical variability of the vortex structure is not considered.}

Figure \ref{fig:para_space}(a) shows $V_{\rm max}$ and $r_{\rm max}$  {for Bu 25 and Bu1 at various streamwise locations}. It can be seen that for both  Bu1 and Bu25, the vortex radii agree reasonably well with the radial location of the least stable $\chi$ (figure \ref{fig:chi_stat}), consistent with theoretical analysis and experimental observation that the edges of vortices are the least stable regions \citep{kloosterziel1991experimental,carnevale1997three,lazar2013inertiala,yim2016stability}. Moreover, AVs in Bu25 have a much larger radii as well as variability during the evolution, compared to Bu1. The AVs in  Bu1, which have greater $V_{\rm max}$ (over twice stronger than  Bu25) and  smaller vortex radii, have the largest average vorticity. 

% \hl{centrifugal instability is active  Often, we implicate centrifugal instability but we need to say something about lack of shear instability or symmetric instability by looking at $Ri_g$! }. 
%
%\jy{jy: are we planning to do it? since there shows almost no centrifugal instability now. }

Figure \ref{fig:para_space}(b) shows the evolution of AVs, on average, in the $Bu_{\rm v}$-$Ro_{\rm v}$ parameter space. The AVs in both Bu25 and Bu1 are characterised by vortex {Rossby} numbers of $O(0.5)$. The stability curves \eqref{lazar} and \eqref{yim} are also plotted for cases Bu1 and Bu25. The left side of a stability curve is the stable region, and vice versa.  
{It can be seen that all AVs in both cases fall on the stable side, and they all tend to evolve to a more stable state (lower $|Ro_{\rm v}|$ and further away from the stability limit). The stability results are in agreement with our observation that there is no apparent sign of instability of AVs except for at $z/h=0.12$ in Bu25, where AVs are still not independently distinguishable from the turbulent near wake. The more conservative determination of cyclo-geostrophic instability in the present wakes utilising \eqref{lazar} and \eqref{yim} as compared to \eqref{gen_ray} also confirms the point of view in \cite{lazar2013inertiala,yim2019stability} that in real geophysical environments, stratification and vertical dissipation will further shrink the range of unstable vertical wavenumbers from the low- and high-wavenumber end, respectively, and lead to a greater range of stability. } 

\section{Mean momentum wake} \label{statistics}

\begin{figure}[pt]\captionsetup[subfloat]{farskip=-12pt,captionskip=1pt}
  \subfloat[]{{\includegraphics[width=.45\linewidth]{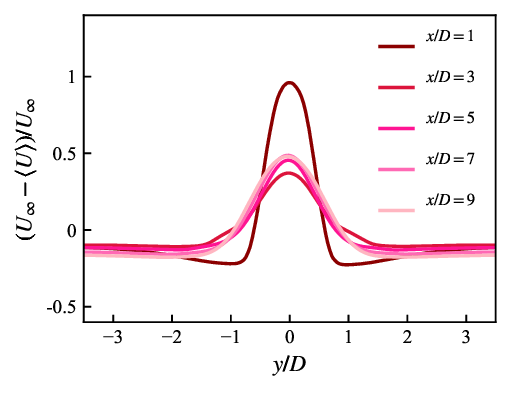}}}% 
  \subfloat[]{{\includegraphics[width=.45\linewidth]{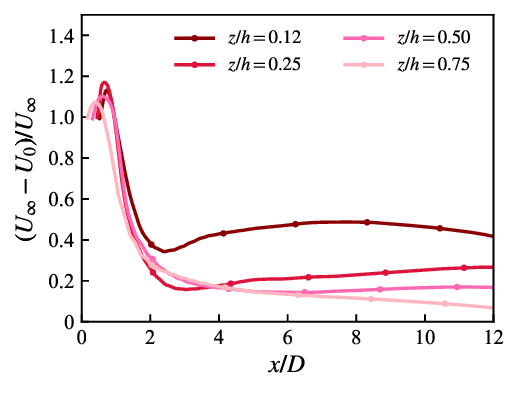}}}% 
  \\ 
  \subfloat[]{{\includegraphics[width=.45\linewidth]{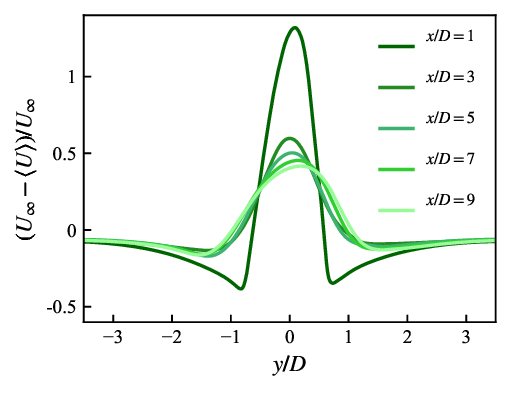}}}% 
  \subfloat[]{{\includegraphics[width=.45\linewidth]{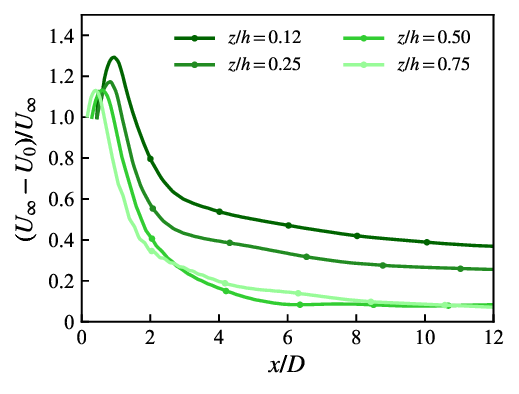}}}% 
  \\ 
  \subfloat[]{{\includegraphics[width=.45\linewidth]{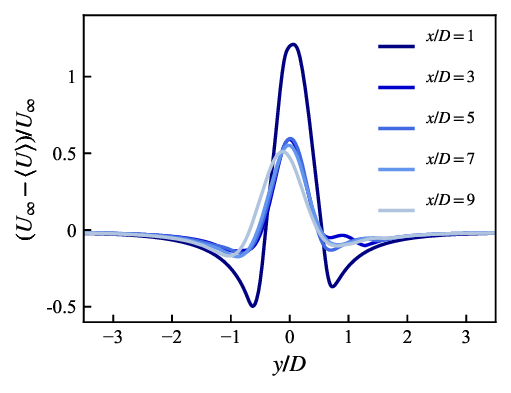}}}% 
  \subfloat[]{{\includegraphics[width=.45\linewidth]{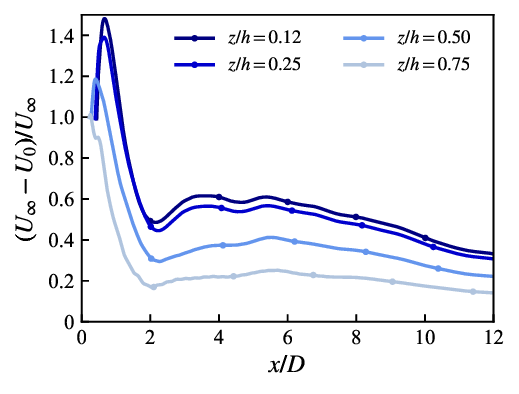}}}% 
  \\ 
  \caption{Time-averaged velocity deficit in the wake for cases BuInf (top row), Bu25 (middle row) and Bu1 (bottom row). Left column (a,c,e) shows transverse profiles of velocity deficit at $z/h=0.12$ and various $x/D=1,3,5,7,9$ (from dark to light). Right column (b, d, f) shows the streamwise evolution of the centerline ($y =0$) deficit at various elevations,  $z/h=0.12, 0.25, 0.50, 0.75$ (from dark to light).}
  %Time-averaged (a,c,e) velocity deficit profiles $\langle U \rangle (x,y,z)$ at $z/h=0.12$ and $x/D=1,3,5,7,9$ (from dark to light), and (b,d,f) centerline velocity deficit $U_0(x,z)$ at $z/h=0.12, 0.25, 0.50, 0.75$ (from dark to light). (a,b) BuInf, (c,d) Bu25, (e,f) Bu1. Different starting points of the lines in (b,d,f) are due to the varying local hill diameters as a function of height.}
  \label{fig:mean_flow} 
\end{figure}

Stratified wakes in engineering applications, e.g. submersibles in the ocean, typically have $Fr \geq O(1)$ and negligible rotation effects. Momentum wakes in these applications are known  to have very different properties compared to their unstratified counterparts, e.g.  a buoyancy-induced slowdown  in   the decay of  mean momentum deficit in the so-called non-equilibrium stage \citep{spedding1997evolution,brucker2010comparative,diamessis2011similarity,de2012simulation}. Stratified wakes, which have been extensively studied for the sphere,   have been investigated recently  for  a blunt body -- a  disk \citep{chongsiripinyo2020decay} and a slender body -- a 6:1 prolate spheroid \citep{ortiz2023high}.  
 In rotating stratified wakes studied here, with the presence of coherent wake vortices  and cyclo-geostrophic balance, the mean flow is further influenced as will be elaborated below. 
%  \Sutanu{ \hl{Remaining lines of the para  are better suited to last section.}  
 Examination of the momentum deficit profiles of this section shows  enhanced persistence of the wake that has implications in oceanography and meteorology. For example, even at $x = 12D$, the  wake deficit behind the near-bottom portion of the hill/seamount is as large as $0.4 U_\infty$. Thus, absent other interacting flow features, bottom roughness or other bathymetry, the wake of a  steep 10 km base-diameter hill  would be  preserved for 120 km. Also, a steep 3D topographic feature would lead to a bottom flow (outside the viscous boundary layer)  with significant shear, for instance,  a difference over obstacle height of 0.2 to 0.4 times $U_\infty$.
  
% The decay of the mean momentum wake is of particular interest in geophysical settings, such as the hill wake. The physical dimension of the hill $D$ could be of $O(10 \, {\rm km})$ large and hence the length of the wake scales up proportionally. The influence region of the obstacle as well as how it is affected by rotation strength would have implications in oceanography and meteorology. 

% Motivation: Coherent structures, vortex stabilities. What are their influnce on the mean wake?
% \begin{figure}[pt]\captionsetup[subfloat]{farskip=-12pt,captionskip=1pt,labelformat=empty}

Figure \ref{fig:mean_flow}(a) shows profiles of mean velocity deficit, $U_d (y) =  U_\infty - \langle U \rangle (y)$, at various  streamwise  locations in the plane $z/h=0.12$ (a,c,e) and figure \ref{fig:mean_flow}(b) shows the streamwise evolution of the centerline deficit velocity ($U_0 = U_d (y =0)$) along various heights (b,d,f). The symbol $\langle \cdot \rangle$ denotes the time average over the duration of data storage listed in table \ref{table1}. 
% \Sutanu{Let's use $U_0$ in line with engineering wake practice.} 
%The centreline mean velocity is $U_0(x,z) = \langle U \rangle (x,0,z)$. 

The non-rotating case BuInf (figure \ref{fig:mean_flow}a) exhibits $U_d(y)$ profiles that are laterally symmetric. The centerline velocity deficit ($U_0 (x)$ in figure \ref{fig:mean_flow}b) initially decreases after the recirculation bubble but  then increases again. This is consistent with the expansion and shrinking of the $\langle \omega_z \rangle (y) $  width  in figure \ref{fig:omgz_mean}(d). In low-$Re$ two-dimensional cylinder wakes,  $U_0 (x)$ has a similar non-monotonic behaviour  \citep{kumar2012origin} and the vorticity profile  width also has a non-monotone variation \citep{barkley2006linear}. On the other hand,  after the recirculation zone, $U_0 (x)$ exhibits monotone decay in three-dimensional unstratified wakes. 
%low past bluff bodies, such as the wake past a circular cylinder \citep{cantwell1983experimental,kravchenko2000numerical} or past a sphere \citep{pal2017direct}, at moderate-to-high Reynolds numbers. 
This similarity of the mean deficit between two-dimensional wakes and non-rotating stratified wakes, where the third dimension is confined by buoyancy, is intriguing.
% implies  potential for  reduced-order modelling of some flow features by  a quasi-twodimensional (Q2D)  {\em ansatz}. 
% Q2D aspect of the present stratified wakes. 

In the rotating cases Bu25 and Bu1, the lateral  symmetry of the mean wake is lost as shown in figure \ref{fig:mean_flow}(c,e). Rotation at $Ro=0.75$ (Bu25) leads to the instability and diffusivity of AVs (commented on previously) and results in a wider wake compared to BuInf.  However, stronger rotation at $Ro=0.15$ (Bu1) creates tall shedding `columns' that are compact and almost `frozen' during evolution, leading to a narrower wake instead. Comparing the centreline deficit (figure \ref{fig:mean_flow}), case Bu1 sustains the highest overall velocity deficit (roughly above $0.2 U_{\infty}$ at all planes) compared to the other two cases, presumably due to the lower diffusivity and higher degree of vortex coherence. In the lower portion of the hill ($z/h =  0.12$) all three cases exhibit  significant $U_0 \approx 0.4 U_\infty$ showing persistence of the near-bottom wake in stratified flow, independent of rotation. 

\section{Concluding remarks} \label{conclusion}

%\jy{jy: The entire section is largely rewritten. The aim was to stand at a higher point and gradually expand the scope of discussion. } 

Wake vortices in stratified flows past an isolated bottom obstacle are studied via LES, at  {moderately high} Reynolds number ($Re_D=10\,000$) and moderately strong stratification ($Fr=0.15$). The rotation Rossby number is varied to include cases representing non-rotating ($Ro=\infty$), submesoscale ($Ro=0.75$), and mesoscale ($Ro=0.15$) wakes, resulting in Burger numbers of $\infty, 25, 1$. In all three cases studied, the wakes present Q2D K{\'a}rm{\'a}n vortex shedding in horizontal planes in the core of the wakes, which is a distinct feature of strongly stratified wakes {at $Fr<O(0.5)$.
}

{In non-rotating wakes at low Froude number, as the vertical overturning motions are constrained, organised motions emerge in the horizontal directions as K{\'a}rm{\'a}n vortices \citep{pal2016regeneration,chongsiripinyo2017vortex}. Stratification is crucial to the formation of the K{\'a}rm{\'a}n vortices at moderately high Reynolds numbers. These dominant horizontal motions were found, instantaneously, to exhibit  coherence in the vertical direction as streamwise-slanted stuctures.  }
%across a vertical length scale numerically larger than $U_{\infty}/N$ 
%but are  slanted in the streamwise direction.
% (such as the `surfboards' in \cite{chongsiripinyo2017vortex}).  }
% \Sutanu{ Jinyuan, the vertical length scale is numerically larger than $U/N$ but, likely, scales as $U/N$ which will become clear in the new simulations with variable $Fr$.} 

{In this study, the vertical coherence of planar K{\'a}rm{\'a}n VS is statistically investigated with SPOD, and the influence of various levels of rotation is included. }
% The statistical spatio-temporal coherence of the K{\'a}rm{\'a}n vortices is analysed with SPOD.
It is found that, at the present stratification ($Fr=0.15$) and regardless of the rotation strength, the shedding frequency is a global constant in each case that is independent of height from $z/h=0.12$ to $z/h=0.75$, which is the core of the wake away from the bottom boundary and the region influenced by the lee waves near the top of the hill.  This implies that, at the present stratification, VS frequency can stay coupled vertically, instead of varying as a function of the local ($z$-dependent) diameter.
{The vertical coupling occurs even when rotation is weak compared to stratification, a result that is inconsistent with the result of \cite{perfect2018vortex} that vortices are vertically decoupled when $Bu^* > 12$ or, converting to our definition, $Bu > 3$. }
% \revv{However, it remains to be seen in future work at what $Fr$ this vertical coherence will be lost and the VS frequency start vary with elevation.}
{However, it is possible that, when stratification is further increased keeping $Bu$ constant, the VS frequency could vary with elevation, which remains to be seen in future work}. 
% \jy{jy: now I suspect that we could still see coupling at $Fr=0.05, Ro=\infty$.}  
% \jy{jy: the conclusion here is that this paper is to be sent to them to referee :)}

{Regarding the influence of rotation on VS,} \cite{boyer1987stratified} measured the VS frequency in a single plane as part of their experimental study of the wake behind a conical obstacle  in linearly stratified rotating flows. The frequency showed little variation in their parametric study that spanned $0.08<Fr<0.28$, $0.06<Ro<0.4$, and three Reynolds numbers $Re_D=380,760,1140$. { The present work confirms the  lack of rotational influence on the VS frequency and adds new information by considering a wider range of rotation strengths, namely $Ro=0.15, 0.75, \infty$, and finding VS  to be at a global height-independent frequency instead of the prior laboratory result  at a single height. Moreover, the present Reynolds number is an order of magnitude larger, implying that the VS frequency is robust to the perturbation of near wake turbulence.}

{On the other hand, the vertical structure of the coherent global modes is significantly affected by strong rotation.} An implication of a global shedding frequency is that the spatial assemblies of SPOD eigenmodes shown in figures \ref{fig:spod}-\ref{fig:spoe} are 3D global modes that optimally represent the vertical enstrophy of the flow. In the non-rotating case BuInf (figure \ref{fig:spod}(a,b)), VS structures are slanted (yet three-dimensionally coherent) `tongues' that tilt to the direction of the flow and form a very shallow angle (steeper near the bottom and shallower above, but at $4^{\circ}$ on average) with the horizontal. { As the rotation increases, the shape of the global modes change from slanted `tongues' to upright `columns'. However, once the coherent structures are formed, their shapes are preserved during the downstream evolution, which is explained by a vortex advection velocity of approximately $0.9U_{\infty}$ that is constant at different heights and in all three cases. }

It is worth noting that despite the turbulence in the  near wake, the flow exhibits overall low-rank behaviour globally owing to the emergence of coherent structures. The low-rankness is two-fold. First, the enstrophy spectra in figure \ref{fig:eigens} show dominant spikes at the VS shedding frequency and its harmonics. Second, the gaps between eigenspectra shown in figure \ref{fig:eigengap} indicate increasingly lower enstrophy in higher-order modes at each frequency. That being said, the large scales of the flow can be well described by a finite set of harmonic modes. This simplicity might encourage future reduced-order modelling of the coherent motion of wake eddies in similar parameter regimes. 
% \revv{Chifan} 

% \st{Besides the macroscopic view of the vortex structures as global modes and how they vary at different $Ro$, a detailed examination of the profiles and evolution of the vortices is conducted. }
{A novel way of tracking vortices automatically in time-resolved snapshots is proposed and applied on the LES database.} Vortex centres (centres of regions of strong $\omega_z$) are extracted with the mean shift algorithm \citep{fukunaga1975estimation,comaniciu2002mean} in each snapshot on 2D horizontal planes. Then the history of vortex centres in time is compiled into graphs that represent evolution trajectories. {The tracking of the vortices has enabled estimation of their advection velocity, quantification of asymmetry between CVs and AVs, and examination of vortex stability properties following their evolution, all in a statistical sense.}

% Finally, conditional statistics are gathered and ensemble averaged on the paths of vortex evolution. 

{The vortex advection velocity, extracted from the time history of vortex centres, is found to be near $0.9 U_{\infty}$ in all three $Bu$ cases, which is quite constant vertically as well. The vertically constant advection velocity explains the preserved vertical orientation of vortex structures during evolution, in all cases. However, it is slightly uneven between cyclones and anticyclones in the rotating cases presumably due to their size difference.}
%    The value of advection velocity also agrees with that in turbulent cylinder wakes \citep{zhou1992convection}, . 

{Regarding  coherent VS structures, the fate of individual vortices is also critical, which is greatly influenced, either being strengthened or weakened, by background rotation. When rotation is strong enough ($Ro=0.15$, case Bu1), VS structures are vertically aligned, reminiscent of Taylor columns, and coherence is enhanced. On the other hand, in conditions favorable for inertial-centrifugal instability, anticyclones could break into turbulence and hence break down the vertical coherence. The adjustment of $\omega_z$ to rotation as well as the cyclo-geostrophic instability of AVs, are hence crucial and analysed by computing the statistics conditioned to the  tracked vortex centres.}  The conditionally averaged vorticity profiles reveal that dynamical processes during the downstream advection of the vortices depend substantially on  rotation. In case BuInf, the distribution of $\omega_z$ is Gaussian-like, with the diffusion-induced downstream increase of size {and decrease of peak vorticity} being the major feature. In case Bu25, the asymmetry between CVs and AVs is most significant among all three cases. {It is worth noting that, when normalized by $f_c$, it is the submesoscale topography with $Bu = 25$ and $Ro = 0.75$ that results in the largest magnitude of  $|\omega_z/f_{\rm c}|$ and not the Bu1 case.}
%and the magnitude of $|\omega_z/f_{\rm c}|$ is also higher than in the other rotating case (Bu1). 
But, in convective units, i.e., $|\omega_z D/U_{\infty}|$,  it is the Bu1 case where CVs and AVs achieve the greatest magnitude of  vorticity   due to the enhanced coherence induced by rotation. Also, both AVs and CVs experience little change during their  evolution 
%even the inertial length scale for geostrophic adjustment $U_{\infty}/f_{\rm c} = RoD$ is smaller in Bu1 than in Bu25. This 
implying  that the vortices in case Bu1 are already in a balanced state. 

Relatively strong anticyclones ({e.g.,} core relative vorticity $\omega_z/f_{\rm c} \approxeq -1.3, -2.4$ for Bu1 and Bu25 respectively, at $z/h=0.12$) are found stable for a considerable distance of advection (figure \ref{fig:vor_ctr_stat}(c,e)). They would be unstable according to the absolute vorticity criterion \eqref{abs_vor}, which implies inertial instability when anticyclonic vorticity is stronger than the Coriolis frequency ($\omega_z/f_{\rm c}<-1$). 
% or marginally unstable at the periphery according to the generalised Reyleigh discriminant \eqref{gen_ray}. However, such loss of stability is not observed during their evolution. 
% To tackle the stability problem of AVs in cases Bu25 and Bu1, various stability criteria are tested. It is found that the absolute vorticity criterion , is not sufficient. 
The inclusion of centrifugal contribution is shown as necessary by examining the generalised Rayleigh criterion \citep{kloosterziel1991experimental,mutabazi1992gap}, which predicts overall stability in the bulk of the AVs and marginal instabilities at the edges, in agreement with observations of stable AVs in the wake. 
% corrects the centrifugal contribution and appears to be good enough, 
Two new recently proposed  criteria \citep{lazar2013inertiala,yim2019stability} considering the effects of  stratification and vertical dissipation are also tested. Both criteria are in terms of the marginal stability curves given by \eqref{lazar} and \eqref{yim} in the parameter space of $Bu_{\rm v}$-$Ro_{\rm v}$, where $Bu_{\rm v}$ and $Ro_{\rm v}$ are the local (vortex) Burger and Rossby numbers specific to the local vorticity profile. Further restricting the range of unstable vertical wavenumbers by including stratification and dissipation, the criteria \eqref{lazar} and \eqref{yim} both determine the AVs in the present work as stable as shown in figure \ref{fig:para_space}(b). The only marginally unstable wake vortices are found in case Bu25 at $z/h=0.12$ (see $x/D=2{\text -}3$ in the last row of figure \ref{fig:omgz_mean}b and figure \ref{fig:chi_stat}a), where the left side of the eddies at downstream location $x/D \sim 4$ is unstable. Further downstream, the region of instability disappears.  

% This instability and later evolution towards stability is identified by the new criterion \eqref{yim} of \cite{yim2019stability}, as shown in figure \ref{fig:para_space}. 

% Besides, statistically, all AVs tend to evolve to a more stable state with an increasing $\chi$ or smaller vortex Rossby number $Ro_{\rm v}$. 

% Our findings suggest that the use of criteria that includes stratification and dissipation such as \eqref{lazar} and \eqref{yim} are essential to the determination of the stability of geophysical eddies. 

Statistically, when the AVs evolve downstream, they tend to approach a more stable state characterised by a larger (more stable) $\chi$ (figure \ref{fig:chi_stat}), a smaller vortex Rossby number $Ro_{\rm v}$ (figure \ref{fig:para_space}a), and a greater distance from the marginal stability curve (figure \ref{fig:para_space}b). 
It is noted that in the present wakes, AVs are observed to be stable in the streamwise extent of vortex tracking (after $x/D \sim 3$). Since both Rossby numbers studied (0.75 and 0.15) are smaller than order unity, a Rossby radius of deformation defined as $U_{\infty}/|f_{\rm c}| = Ro D$ which can be regarded as a distance required for cyclo-geostrophic adjustment, will be smaller than $3D$ downstream. It is possible that at larger Rossby numbers, more unstable AVs will be observed further than $x/D>3$ where the vortex tracking and stability determination are not obscured by near-wake turbulence. {Overall, the result of stability AVs downstream is consistent with the prolonged coherent motions in cases Bu25 and Bu1, and the applicability of the criteria in \cite{lazar2013inertiala,yim2019stability} is demonstrated in the simulation of geophysical topographic wakes, with a massive ensemble of samples for statistics.}  

In terms of future work, it would be useful to study wake vortices in other  parameter regimes   with non-hydrostatic simulations. Cases at lower $Fr$ and a wide range of $Ro$ are of particular interest with respect to the variation of vortex shedding frequency. Submesoscale instabilities at  high $Ro$ and high $Re$ are possible and need investigation. Near-wake turbulence and mixing are also important follow-up topics in the context of the broader theme of ocean turbulence and mixing. Theoretical global stability analyses of stratified wakes would also provide a more complete picture. 

\backsection[Funding]{We are pleased to acknowledge  support by ONR grant N00014-22-1-2024.}
\backsection[Declaration of interest]{The authors report no conflict of interest.}
\backsection[Author ORCIDs]{\\ 
\noindent J. Liu \url{https://orcid.org/0000-0003-4133-0930} \\
\noindent P. Puthan \url{https://orcid.org/0000-0003-2690-0560} \\
\noindent S. Sarkar \url{https://orcid.org/0000-0002-9006-3173}}

%%%%%%%%%%%%%%%%%%%%%%%%%%%%%%%%%%%%%%%%%%%%%%%%%%%%%%
\appendix 
%%%%%%%%%%%%%%%%%%%%%%%%%%%%%%%%%%%%%%%%%%%%%%%%%%%%%%
\section{A vortex tracking method for time-resolved databases} \label{app1}

% \subsection{Methodology}
Mean shift \citep{fukunaga1975estimation,comaniciu2002mean} is a widely used method for pattern recognition in data analysis. It identifies centroids of condensed data points and then segments the data according to the centroids they belong to. This method can be applied to physical science as a means of data clustering, where some shared properties are expected for data points in the same cluster. It is unsupervised in the sense that either the number of clusters is required or the shape of clusters is prescribed. The basic idea is to move a provisional centroid iteratively toward the local maximum of the population density, hence it is also called a density-based method. 

Similar to the implementation in \cite{gong2015interactive}, the mean shift algorithm is summarised as follows: 
\begin{enumerate}
  \item[\em Step 1.] Randomly select an initial seed for the $i$-th centroid $V_i^{(0)}$ from all unclustered points,
  or use the centroid computed in the $n$-th iteration $V_i^{(n)}$.  
  \item[\em Step 2.] Compute the centre of geometry (the so-called mean, denoted as $V_i^{(n+1)}$) of the data points that fall in the open ball $\mathcal{B}(V_i^{(n)},r_{\rm BW})$, where $V_i^{(n)}$ is the ball centre and $r_{\rm BW}$ is the radius. % Denote this new centroid as . 
  \item[\em Step 3.] If the Euclidean distance $|V_i^{(n+1)}-V_i^{(n)}|$ between $V_i^{(n+1)}$ and $V_i^{(n)}$ is below the tolerance, accept $V_i^{(n+1)}$ as $V_i$. Otherwise, use $V_i^{(n+1)}$ as the new seed to restart {\em Step 1}. 
  \item[\em Step 4.] Compare the Euclidean distance of the centroid $V_i$, to all existing centroids $\{V_l\}_{l=1}^{i-1}$ from previous iterations, and if $\exists j, \, {\rm s.t.} \, |V_i-V_j|<1/2 r_{\rm BW} (0 < j < i)$, merge $V_i$ and $V_j$ and label their mean as $V_j$. Run through the distance check in {\em Step 4} for $V_j$ in the set $\{V_l\}_{l=1, l \ne j}^{i-1}$. 
  \item[\em Step 5.] Check if there are unvisited points. If yes, start over from {\em Step 1}; otherwise, terminate.  
\end{enumerate} 

\begin{figure}[tp]
  \includegraphics[width=.5\linewidth]{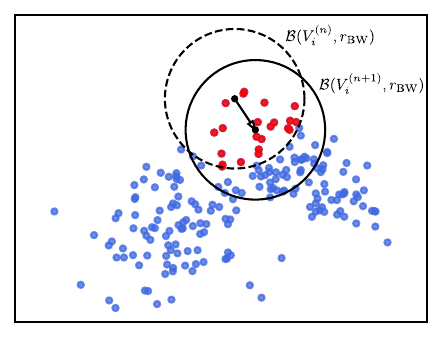}% 
  \caption{Illustration of the principle of the mean shift algorithm. Data points are randomized and are for illustration purposes only. The symbol $\mathcal{B}(V_i^{(n)},r_{\rm BW})$ denotes an open ball with its centre at $V_i^{(n)}$ and radius $r_{\rm BW}$, where $V_i^{(n)}$ is the $n$-th iteration of the $i$-th centroid, and $r_{\rm BW}$ is the half bandwidth. The centroid $V_i^{(n+1)}$ represents the geometric centre of all data points enclosed by the ball $\mathcal{B}(V_i^{(n)},r_{\rm BW})$ shown in red. The arrow denotes the shift of the mean (centroid). }
  \label{fig:mean_shift} 
\end{figure}

% Mean shift is built on kernel density estimation (KDE), which represents discrete data with known kernel functions and estimates the density distribution of the data. 

% An automated method to process all $N_t=4000$ snapshots and extract the trajectories of vortex evolution. 

% For graph, cite \cite{lozano2014time}.

% In this method, the bandwidth $d_{\rm BW}$ needs to be selected based on physical understanding of the data, and will influence the interpretation of the results. 
The core steps 2-3 are illustrated in figure \ref{fig:mean_shift}, where the shift of the mean is indicated by an arrow. Based on the mean shift algorithm, the vortex centre identification and tracking process utilised in this paper is summarised as follows: 

\begin{enumerate}
  \item[(1)] Mask the vorticity field. Convert each 2D vorticity field at a certain height $\omega_z(x,y,t;z)$ into a binary field, with ones denoting points with $\omega_z>\alpha$ (if identifying positive vortex centres) or $\omega_z<-\alpha$ (if identifying negative vortex centres), and zeros denoting the rest. Here a positive constant $\alpha(z)$ is a threshold individually selected for each horizontal plane to disconnect vortices from each other. We note that the hill wake is inhomogeneous in all three spatial directions and a global constant $\alpha$ does not apply. 
  \item[(2)] Apply the mean-shift algorithm to identify (usually a handful of in our flow) centres in each snapshot. Use $V_{i,k}$ to label the $i$-th vortex centre ($1 \le i\le n_{\rm max}$, where $n_{\rm max}$ is the maximum number of centres allowed) in the $k$-th snapshot ($1 \le k \le N_t$). The half bandwidth $r_{\rm BW}(z)$ needs to be selected as a parameter individually for each plane, and is chosen to be roughly 1.5 times the radius of a vortex, which is also smaller than the separation between two same-sign vortices. 
  \item[(3)] Construct the graphs of vortex centre trajectories, with each centre $V_{i,k}$ ($1\le i \le n_{\rm max},1 \le k \le N_t$) being a node. Only the connections (edges) between two nodes from two consecutive snapshots are considered, with the connection weights being the Euclidean distance between these two nodes  {
  % \begin{equation}
    $\Delta x_c=|x_c(V_{i,k})-x_c(V_{j,k+1})|$, 
  % \end{equation}
  where $x_c$ stands for the streamwise location of the centre}.  {Two nodes are considered to belong to the history of the same vortex} (and hence belong to the same subgraph) if $\Delta x_c < 1.5 U_{\infty} \Delta t$, where $\Delta t$ is the time elapsed between these two snapshots. All connection weights that {do not} satisfy this restriction will be set to zero. The choice of  {the separation distance $1.5 U_{\infty} \Delta t$ is meant to be inclusive since it is unlikely for the vortex centres to travel much faster than the background flow. On the other hand, this distance is small enough compared to the distance between two distinct vortices}. After doing so, it is almost ensured that each centre will only have one forward and one backward connection in time. Each self-connected subgraph represents a vortex evolution trajectory within the domain. 
\end{enumerate}

{It is noted that the identification and tracking method described above has limitations, such as requiring user input of a constant radius of searching, which is {\em a priori}  knowledge about the physical system and was chosen to be  around 1.5 times the radius of K{\'a}rm{\'a}n vortices in each plane. Hence, this method may not work in more complicated situations} 
% and could fail in more complicated flows}
% Developing a structure-tracking algorithm is not the focus of this work. We are aware of the limitations that it requires user inputs as prior knowledge and could fail in more complicated flows, 
such as in {flows that have a wide range of scales of vortices, or} in processes involving vortex merging or splitting where two same-sign vortices can get too close. But for {the present} vortex wakes and other similar vortical flows where the {organisation and evolution of vortices} is clear, {the computer-aided identification and tracking scheme} is shown to be useful and easy to implement. Owing to the continuing advancement of computer power and experimental techniques, time-resolved databases are becoming more available, where our snapshot-based tracking method {may facilitate the analysis of coherent structures} in other types of flows and will therefore be of broader interest. 

%%%%%%%%%%%%%%%%%%%%%%%%%%%%%%%%%%%%%%%%%%%%%%%%%%%%%%
%
%  Moved to thesis 

\section{{Analysis of grid and domain sensitivity}} \label{app2} 

In the numerical simulations of bluff body wakes, domain confinement in the transverse direction ($y$) has been shown to have an influence on the instability of low Reynolds number wakes by various approaches. \cite{juniper2006effect} performed a linear normal-mode stability analysis of a top hat velocity profile that mimics the initial one-dimensional velocity profiles ($U(y)$) of a 2D wake and found that the absolute instability can be over-predicted when the domain in the transverse direction is confined. \cite{biancofiore2011influence,biancofiore2012influence} conducted numerical simulations and verified at various Reynolds numbers from laminar to turbulent that, when confined by two slip walls, the wake created by a top hat velocity profile could present different instability modes at different confinement ratios. For the global mode specifically corresponding to the K{\'a}rm{\'a}n VS, \cite{kumar2006effect} showed using a biglobal linear stability analysis of the mean flow of a 2D cylinder wake that $St_{\rm VS}$ is over-predicted near the onset of K{\'a}rm{\'a}n VS (at $Re_D \approxeq 47$), which can be corrected as an effect of blockage that increases the effective Reynolds number. 

In the present study, the effect of finite transverse domain is taken into consideration when conducting the LES. Periodic boundary conditions are used in the transverse direction, which results in less restriction on the horizontal flapping motion of the wake than no-penetration walls. We also note that the confinement effect is eased in 3D wakes as compared to 2D wakes. Furthermore, the present wakes are at $Re_D=10\, 000$, and they feature saturated values of $St_{\rm VS}$ in the $St$-$Re$ correlation of \cite{williamson1998series} at the $Re \rightarrow \infty$ limit. This fact implies that the $St_{\rm VS}$ has entered a regime with less sensitivity to the confinement compared to transitional wakes. In simulations BuInf, Bu25, and Bu1, the highest chance of wake confinement is at the base of the hill where  $L_y/D=8$, a value which is shown to result in  low confinement by  \cite{juniper2006effect} and \cite{biancofiore2011influence}. Nevertheless, to further evaluate the effects of confinement and to exclude its influence on the vortex dynamics studied here, three auxiliary simulations at $Fr=0.15, Ro=\infty,$ and $Re=10\,000$ are performed. The sensitivity of the mean velocity deficit as well as the VS frequency to transverse domain length ($L_y$) and grid resolution ($\Delta y$) is examined. 

Table \ref{tablea2} provides the details of the simulations. Case BuInf is the production simulation. The other three cases are also named BuInf, but with appended `G' and `D' to denote the purpose of  grid- and domain-sensitivity analysis, respectively. In case BuInfG, $L_y$ is kept the same while $\Delta y$ is doubled ($N_y$ halved). In cases BuInfD1 and BuInfD2, the $y$-resolution is similar to BuInfG, but  the  domain widths are increased to $L_y=12$ and $L_y=15$, respectively. For the auxiliary simulations, a period of at least two flow-throughs is chosen before data are collected to avoid contamination by the initial transient. The time span for all statistics is about six flow-throughs, as listed in table \ref{tablea2}.

\begin{table}[t!]
  \caption{Simulation parameters for grid- and domain-independence study. Case BuInf is the same as the one in table \ref{table1}, while the appended `G' denotes grid independence case and `D1', `D2' denote two domain independence cases. The transverse span of the computational domains is $L_y$. }
  \begin{tabular}{l c c c c c c c}
    % \hline 
    \toprule
    Case &$Bu$&$Ro$ & $Fr$                  & $(N_x,N_y,N_z)$ & $L_y/D$ & $N_t$ & $T U_{\infty}/D$  \\ 
    % \hline 
    \midrule 
    BuInf   & \multirow{4}{*}{$\infty$}   & \multirow{4}{*}{$\infty$}  & \multirow{4}{*}{0.15}                     & (1536,1280,320) & $[-4,4]$ &  4000 & 295  \\ 
    BuInfG  & &  &          & (1536,768,320)  & $[-4,4]$       & 1500 & 120  \\
    BuInfD1 & &  &          & (1536,1152,320) & $[-6,6]$       & 1500 & 122   \\
    BuInfD2 & &  &          & (1536,1536,320) & $[-7.5,7.5]$   & 1500 & 116   \\
    % \hline 
    \bottomrule 
  \end{tabular} 
  \label{tablea2}
\end{table}

\begin{figure}[pt]\captionsetup[subfloat]{farskip=-12pt,captionskip=1pt}
  % \subfloat[]{{\includegraphics[width=.45\linewidth]{./figs/ux_i0020.eps}}}% 
  % \subfloat[]{{\includegraphics[width=.45\linewidth]{./figs/ux_i0038.eps}}}% 
  % \\ 
  \subfloat[]{{\includegraphics[width=.45\linewidth]{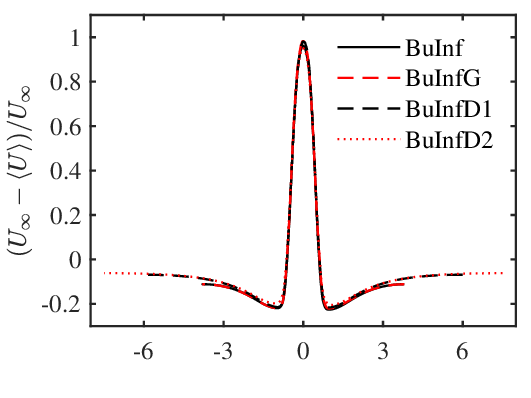}}}% 
  \subfloat[]{{\includegraphics[width=.45\linewidth]{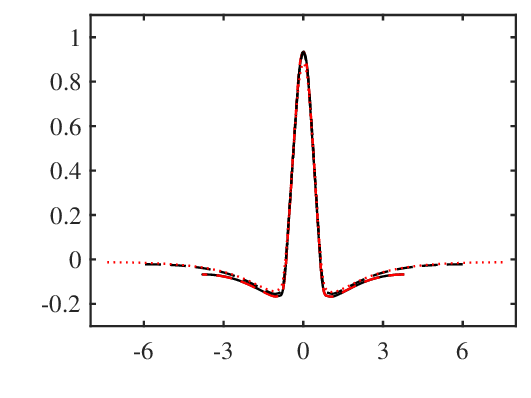}}}% 
  \\ 
  \subfloat[]{{\includegraphics[width=.45\linewidth]{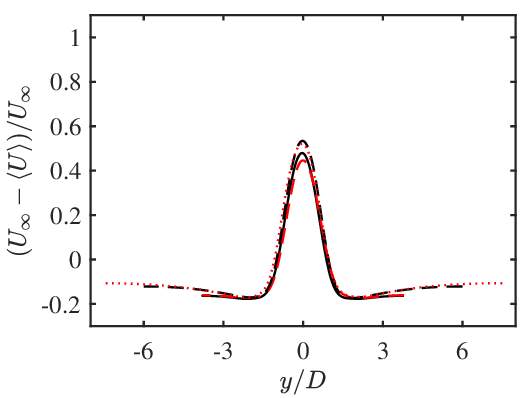}}}% 
  \subfloat[]{{\includegraphics[width=.45\linewidth]{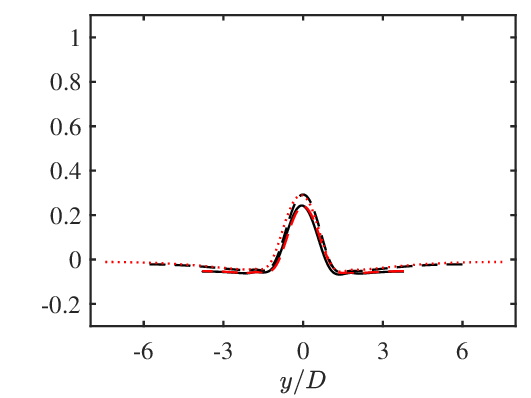}}}% 
  \\ 
  \caption{Mean velocity deficit profiles. Left column corresponds to  $z/h=0.12$; right column to $z/h=0.25$; top row corresponds to $x/D=1$; bottom row to $x/D=9$. }
  %Time-averaged (a,c,e) velocity deficit profiles $\langle U \rangle (x,y,z)$ at $z/h=0.12$ and $x/D=1,3,5,7,9$ (from dark to light), and (b,d,f) centerline velocity deficit $U_0(x,z)$ at $z/h=0.12, 0.25, 0.50, 0.75$ (from dark to light). (a,b) BuInf, (c,d) Bu25, (e,f) Bu1. Different starting points of the lines in (b,d,f) are due to the varying local hill diameters as a function of height.}
  \label{fig:mean_gd} 
\end{figure}

Figure \ref{fig:mean_gd} shows the mean velocity deficit at two horizontal planes ($z/h=0.12, 0.25$) and at two streamwise locations ($x/D=1,9$), as functions of $y$. For streamwise location $x/D=1$ in the near wake, the difference between cases is negligible at the centre of the wake, and is larger when the lateral edge of the domain is approached. As for the streamwise location $x/D=9$ in the intermediate wake, the centreline velocity deficit shows a small difference between cases BuInf and BuInfG with a narrower domain, and the other two, BuInfD1 and BuInfD2, with larger domains. This is due to the growth of the wake in $x$ direction and the increased influence of transverse confinement as compared to the wake width. But in all, the wake deficit profiles show little dependence on the grid resolution, and case BuInfD1 ($L_y=12$) is showing a convergence towards BuInfD2 ($L_y=15$), so we conclude that 
for the accuracy of the wake deficit in the near-to-intermediate wake, 
the present simulation (BuInf) is adequate, but the domain width and the resolution of BuInfD1 can provide a slight improvement. When simulating  the same wakes further downstream, wider domains are necessary. 
  
\begin{figure}[htp]
  \includegraphics[width=1\textwidth]{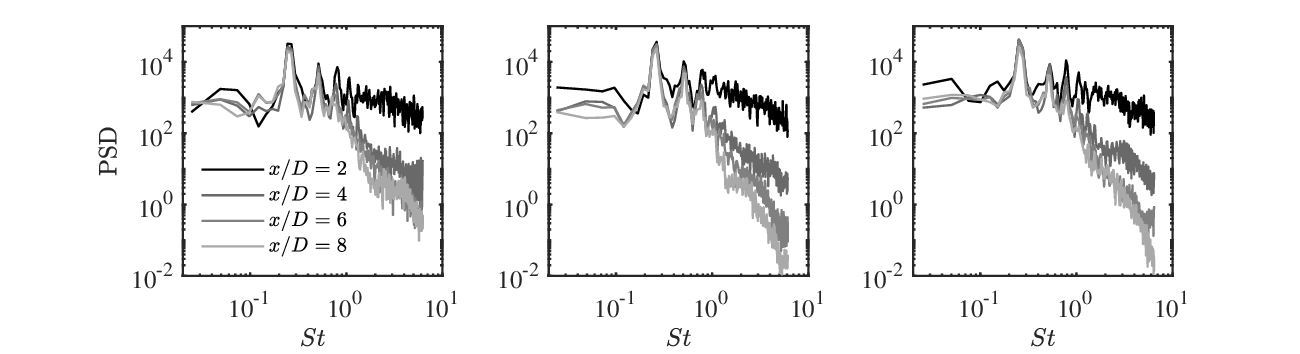}
  \caption{The power spectral density of $\omega_z$ at $z/h=0.12,y/D=0.6$ and $x/D=2,4,6,8$, for cases (a) BuInfG, (b) BuInfD1, and (c) BuInfD2. The VS Strouhal's in (a-c) are $St_{\rm VS}=0.245,0.265,0.252$.} 
  \label{fig:app2_a} 
\end{figure}

% The length of recirculation bubble is an important measure of the quality of wake simulations. 
In addition to the mean statistics, the most important measure of the unsteadiness in the present wakes is the VS shedding frequency. For cases BuInfG, BuInfD1, and BuInfD2, $St_{\rm VS}$ is obtained from single-point Fourier spectra. The time-series of $\omega_z$ are obtained from various downstream locations in the wake ($x/D=2,4,6,8$) and at $z/h=0.12,y/D=0.6$.  Following a similar procedure of SPOD described in section \ref{spod-theory}, the signal is interpolated with PCHIP to a uniform time spacing, and chopped into overlapped blocks of length $N_{\rm FFT}=512$. Hamming-windowed FFT is performed on each block and the power spectral density (PSD) is ensemble-averaged as in a standard Welch's method. The sampled PSD is shown in  figure \ref{fig:app2_a}. 

In the three cases shown in figure \ref{fig:app2_a}, the magnitudes of the VS frequency are consistent with the SPOD leading frequency in BuInf ($St_{\rm VS}=0.264$), and the fluctuations are within the FFT frequency resolution  ($\Delta St \approxeq 0.025$ for all cases). The VS frequency in the other planes ($z/h=0.25,0.50, 0.75$) are the same as in plane $z/h=0.12$, for each case (not shown). It is thus verified that the global VS frequency  as well as its vertical coupling are also not sensitive to the chosen values of $L_y$ and $\Delta y$.

% The PSD is calculated from a windowed FFT with the Welch's method, with $N_{\rm FFT}=512$ same as used in SPOD. Before PSD is computed, the time-series is interpolated in time to obtain even spacing. 

%%%%%%%%%%%%%%%%%%%%%%%%%%%%%%%%%%%%%%%%%%%%%%%%%%%%%%
\bibliographystyle{jfm}
\bibliography{jjl} 

\end{document}